\documentclass[10pt, final]{IEEEtran}
\usepackage{amsthm}

\usepackage{amssymb}
\usepackage{amsmath}
\usepackage{bbm}
\usepackage[linesnumbered,ruled,vlined]{algorithm2e}
\usepackage{mathrsfs}
\usepackage{cite}
\usepackage{bm,epsfig,amsthm,url}
\usepackage{indentfirst}
\usepackage{amsfonts}
\usepackage{epstopdf}
\usepackage{cases}
\usepackage{cuted}
\usepackage[caption=false,font=scriptsize]{subfig}
\usepackage{xcolor}
\usepackage{mathtools}

\makeatletter
\renewcommand{\maketag@@@}[1]{\hbox{\m@th\normalsize\normalfont#1}}%
\makeatother
\usepackage{hyperref}
\usepackage{stfloats}
\usepackage{algpseudocode}
\usepackage{booktabs}
\usepackage{hyperref}
\usepackage{threeparttable}

\newtheoremstyle{mystyle}{}{}{}{}{}{: }{0pt}{\indent \it{\thmname{#1}\thmnumber{ #2}\thmnote{#3}}}
\theoremstyle{mystyle}

\newtheorem{Proposition}{Proposition}

\usepackage[letterpaper, left=0.68in, right=0.68in, bottom=1in, top=0.72in]{geometry}
\def\BibTeX{{\rm B\kern-.05em{\sc i\kern-.025em b}\kern-.08em
    T\kern-.1667em\lower.7ex\hbox{E}\kern-.125emX}}
\allowdisplaybreaks[4]

\begin{document}
\title{Mode Switching for RDARS-Aided ISAC Systems: From Optimization to Deep Unfolding}
\author{Chengwang~Ji, Haiquan~Lu, Qiaoyan~Peng, Jintao~Wang, Feifei~Gao,~and~Shaodan~Ma
\thanks{This article was presented in part at the IEEE ICC 2026 Workshops \cite{ji_ICC_2026}.

C. Ji, Q. Peng, J. Wang, and S. Ma are with the State Key Laboratory of Internet of Things for Smart City and the Department of Electrical and Computer Engineering, University of Macau, Macao SAR, China (e-mails: ji.chengwang@connect.um.edu.mo; qiaoyan.peng@connect.um.edu.mo; wang.jintao@connect.um.edu.mo; shaodanma@um.edu.mo).

H. Lu is with the School of Electronic and Optical Engineering, Nanjing University of Science and Technology, Nanjing 210094, China (e-mail: haiquanlu@njust.edu.cn).

F. Gao is with Department of Automation, Tsinghua University, Beijing 100084, China (email: feifeigao@ieee.org).}
\vspace{-15pt}
}
\maketitle
\begin{abstract}
Reconfigurable distributed antennas and reflecting surface (RDARS) has recently emerged as a promising architecture for integrated sensing and communication (ISAC), owing to its flexible element-wise mode switching between connection and reflection modes. 
In this paper, to fully reap the benefits of mode configuration, muting elements that can absorb the incident energy are introduced into RDARS-aided ISAC systems to mitigate multi-user interference (MUI) and enhance sensing performance. 
To draw useful insights, we first investigate the special cases of single-UE communication, single-target sensing, and two-UE communication to reveal the importance of muting elements. 
Specifically, the maximum communication and sensing signal-to-noise ratio (SNR), and the signal-to-interference-plus-noise ratio (SINR) expressions are respectively derived for the three cases, together with the optimal number of muting elements for explicitly characterizing the tradeoff between reflection gain loss and MUI suppression. 
Next, we consider the joint waveform and tri-mode switching design for RDARS-aided ISAC systems, where an alternating optimization (AO)-based penalty dual decomposition (APDD) algorithm is proposed to solve the mixed-integer nonlinear programming (MINLP) problem. 
Furthermore, a model-driven APDD-Net is developed by deeply unfolding the APDD iterations into a layer-wise architecture, where key parameters are learned to reduce the computational complexity and accelerate convergence. 
Simulation results verify the theoretical findings on the muting gain and demonstrate that the proposed APDD-Net achieves a better tradeoff between communication and sensing performance compared with benchmark schemes.
\end{abstract}


{\section{Introduction}}
Reconfigurable distributed antennas and reflecting surface (RDARS) has been proposed as a novel architecture recently, which integrates distributed antennas and cost-effective passive elements \cite{ji_ICC_2026, ChengzhiMa_ANewArchi, Wang_RDARS}. Specifically, RDARS consists of a large number of reconfigurable elements that can dynamically switch between two modes via the switching network, i.e., connection mode and reflection mode \cite{Wang_RDARS}. The element working in connection mode acts as a conventional active antenna capable of transmitting or receiving signals, while for the reflection mode, the element passively reflects the impinging signal, as in the reconfigurable intelligent surface (RIS).
Besides, only a limited number of radio frequency (RF) chains are connected to the RDARS via cables or RF-over-fiber (RFoF) links for supporting the dynamic mode switching. 
Such an architecture simultaneously brings the reflection and distribution gains, which have been validated through both theoretical analysis and experimental demonstrations in \cite{ChengzhiMa_ANewArchi}. 

Driven by the selection gain brought by the flexible switching of working modes, considerable research efforts have been devoted to RDARS-aided communication \cite{Wang_RDARS, ji2026joint, lu2024tutorial, xue_survey, ji2025model} and integrated sensing and communication (ISAC) systems \cite{jintao_ISAC_RDARS, zhang_RDARS}. 
In \cite{Wang_RDARS}, 
the binary constraint related to mode switching was equivalently transformed into a more tractable formulation and solved by the majorization-minimization (MM) technique.
To reduce the complexity of mode selection, the connected elements were arranged as a uniform sparse array in \cite{ji2026joint}, and thus the mode selection simplifies to the sparsity level design.
In \cite{ji2025model}, a model-driven deep learning enhanced mode selection was proposed to accelerate the convergence.
For ISAC systems, in \cite{jintao_ISAC_RDARS}, a RDARS-aided ISAC prototype was developed to localize a mobile user without degrading the communication performance. 
However, the positions and number of the connected elements were fixed, which limits the achievable mode selection gain. 
In \cite{zhang_RDARS}, the radar output signal-to-noise ratio (SNR) was maximized via the joint optimization of beamforming and mode selection, where the mode selection subproblem was reformulated as a sorting problem.



Note that beyond the conventional reflection and transmission modes, research exploring the diverse functionalities of RIS elements has gained significant momentum \cite{liuSTARSimultaneousTransmission2021, nguyenDecisionDirectedHybridRIS2024, peng_semiRIS, ntontin2024perpetual, huang2024hybrid, xieReflectNotReflect2023, chen_active_RIS, shaoTargetMountedIntelligentReflecting2024, jammer_Tabe_mute}.
For example, the signal is refracted toward the opposite side of the surface via the elements in simultaneously transmitting and reflecting (STAR)-RIS \cite{liuSTARSimultaneousTransmission2021}. The sensors integrated in the semi-passive RIS can sample the incident signal via active RF receivers for channel estimation or sensing \cite{nguyenDecisionDirectedHybridRIS2024, peng_semiRIS}. Moreover, the energy is harvested by the absorptive elements to further improve the energy efficiency in \cite{ntontin2024perpetual}. Specifically, the RF energy absorbed by subsets of absorptive elements at RIS is combined and converted into direct current (DC) power via rectifiers, where the harvested energy is utilized to power the application-specific integrated circuits controlling passive elements \cite{ntontin2024perpetual}.
On the other hand, the incident signal was directly dissipated via the muting elements \cite{xieReflectNotReflect2023, shaoTargetMountedIntelligentReflecting2024, jammer_Tabe_mute}. 
In \cite{jammer_Tabe_mute}, the multi-user interference (MUI) was mitigated via the muting elements, where 
the absorption mode can be achieved via the on-off control mechanism \cite{xieReflectNotReflect2023}.
Thanks to the advantages of muting elements,
integrating the muting mode into the RDARS architecture is promising by providing additional flexibility for mitigating MUI and enhancing sensing performance in ISAC systems. 
Nevertheless, existing studies on the muting mode mainly focus on RIS architectures, and the impact of muting elements on MUI suppression and sensing performance has not been investigated.




Recently, ISAC has attracted widespread research interest as a promising technology for the sixth-generation (6G) wireless networks \cite{liufan_waveform, lu_UAV, wang2024robust, SHI2020107530, shi2024constant, afshani2024combining, guo2024retrodirective}. Specifically, 
the weighted sum of MUI and waveform discrepancy problem that aims to generate a dual-functional (DF) waveform was investigated to strike a balance between the sensing and communication performance \cite{liufan_waveform, wang2024robust, SHI2020107530, shi2024constant}.
In \cite{liufan_waveform}, the designed waveform significantly improves the communication performance by allowing a slight radar performance loss.
The work \cite{SHI2020107530} proposed an iterative algorithm based on the block successive upper-bound minimization framework to minimize MUI while ensuring that the synthesized transmit beam pattern closely approaches the desired one.
The transmit DF waveform was optimized in \cite{shi2024constant} by using the penalty term-based Riemannian conjugate gradient method. 
However, joint optimization of the DF waveform, mode selection, and beamforming for balancing communication and sensing performance incurs high complexity.

To solve the above problem, the deep learning-based methods have been widely investigated in existing ISAC and RIS-aided systems \cite{Jiangpeng_ISAC_SLP_DL, saikia2024hybrid, jiang_ISAC_Net_2024, Miguel_hardware_MD, yuan2024low, jiangpeng_ISAC_MD_2025, lin2024single,  zhang2025joint_deep_unfolding}. 
In \cite{Jiangpeng_ISAC_SLP_DL}, a lightweight network was proposed to obtain the symbol-level precoding vector. 
A hybrid deep reinforcement learning (DRL) algorithm was proposed in \cite{saikia2024hybrid}, where the high-dimensional decision-making problem was reformulated as a Markov decision process.
While the DRL and unsupervised learning neural network (NN) demonstrate the potential performance gains, they may demand a high-quality dataset, extensive fine-tuning, and lack interpretability. 
Different from data-driven methods, deeply unfolding NNs are utilized to achieve fast convergence, and enhance overall performance by incorporating expert knowledge into conventional learning networks in ISAC systems, which are regarded as model-driven methods \cite{jiang_ISAC_Net_2024, Miguel_hardware_MD, jiangpeng_ISAC_MD_2025, zhang2025joint_deep_unfolding, he2024signal}. 
Specifically, the model-driven ISAC-NET in \cite{jiang_ISAC_Net_2024} unfolded a conventional alternating optimization algorithm into a deep learning architecture, where only a few hyperparameters, e.g., step sizes, were trained, yet achieving the low NMSE and bit error rate (BER). By considering hardware impairments, the authors in \cite{Miguel_hardware_MD} introduced a framework where a learnable matrix and vector were optimized via dictionary learning and impairment learning, respectively, thereby reducing the overall complexity. In \cite{jiangpeng_ISAC_MD_2025}, the Powell–Hestenes–Rockafellar transformation was applied to address the highly nonlinear direction-of-arrival (DoA) estimation problem. 
Furthermore, an approximate matrix inversion method was incorporated to further reduce computational complexity, where the learnable parameters associated with the approximation were optimized through a deep unfolding network \cite{zhang2025joint_deep_unfolding}.

{Motivated by the above considerations, and to fully unleash the potential of RDARS in ISAC systems, we extend existing RDARS elements to the tri-mode elements that switch among the connection, reflection, and muting modes, so as to enhance system performance by flexibly reshaping the beam pattern and mitigating MUI. However, such a configuration introduces several fundamental challenges for the considered systems.}
First, it is challenging to formulate a unified optimization problem that balances sensing and communication performance due to the strong coupling among system variables. In particular, the DF waveform is designed to simultaneously serve communication users and sensing tasks, leading to an intrinsic tradeoff between communication and sensing performance.
Second, the impact of muting elements on communication and sensing performance and the corresponding gain remains unclear.
Third, it is difficult to design an efficient algorithm to solve the optimization problem involving the joint optimization of the waveform, passive beamforming, and mode switching matrices. In addition, the optimization of dynamic mode configurations leads to high computational complexity.

To tackle the above challenges, in this paper, we consider a RDARS-assisted multi-user ISAC system, where the waveform, beamforming design, and flexible connection and muting mode switching matrices are jointly optimized to enhance the system performance. To draw useful insights, the special cases of single-UE, single-target, and two-UE communication are investigated, where the optimal SNR/SINR and the optimal number of muting elements are derived. Moreover, the additional muting gain brought by muting elements is verified. 
Furthermore, a deep unfolding network based on the AO-based penalty dual decomposition (APDD) algorithm is proposed to accelerate the convergence and improve the system performance.
The main contributions of this paper are summarized as follows: 
\begin{itemize}
\item First, the muting mode is introduced to RDARS-aided ISAC systems. To reveal the \textit{muting gain} brought by muting modes, we first investigate several special cases, including single-UE, single-target, and two-UE scenarios, and it is shown that muting elements can yield a SINR improvement in the multi-user case. In particular, for the two-UE case, we derive a closed-form expression for the optimal number of muting elements, and explicitly quantify the trade-off between MUI suppression and reflection gain.

\item Second, we formulate a unified optimization framework to minimize the weighted sum of MUI energy and waveform discrepancy with respect to the FMCW reference waveform. To address the resulting non-convex mixed-integer nonlinear programming (MINLP) problem, we propose an efficient APDD algorithm. Specifically, the original problem is decomposed into tractable subproblems, where a closed-form solution is derived for waveform design, and power iteration algorithm is employed for passive beamforming. Moreover, the subproblems of mode switching are reduced to sorting problems via the MM method.

\item Third, a model-driven deep learning-based APDD-Net is proposed by unfolding the APDD algorithm into a trainable architecture, where each layer of the network is associated with the iterations of the APDD algorithm.
In particular, the closed-form expression of the optimal waveform is derived, which removes the need for iterative updates. Furthermore, the matrix inversion is replaced by an approximation with trainable matrices, thus reducing the complexity. The regularization term for passive beamforming is also learned to guarantee the positive definiteness of the involved matrix, while the penalty term associated with binary mode selection is adaptively optimized in each layer, so as to improve the convergence and performance.

\item Lastly, simulation results demonstrate the advantages of the DF waveform design and the effectiveness of the proposed algorithms. It is shown that the APDD-Net achieves a balance between the minimization of MUI energy and waveform discrepancy. Moreover, it is found that the muting elements yield an additional muting gain, in terms of waveform discrepancy reduction and MUI suppression.
\end{itemize}

The rest of this paper is organized as follows. Section \ref{sec: system model} introduces the system model and formulates the MINLP problem. In Section \ref{sec: Special Cases}, the single-UE, single-target, and two-UE cases are analyzed. In Section \ref{sec: APDD}, the joint design of DF waveform, passive beamforming for RDARS passive elements, connection and muting mode switching matrices at the RDARS are studied. 
Section \ref{sec: APDD-Net} presents the proposed deeply unfolding network, i.e., APDD-Net. 
Simulation results are presented in Section \ref{sec: simulation}, and this paper is concluded in Section \ref{sec: conclusion}.

\textit{Notations:} For a complex vector $\bf x$, $x_i$ represents the $i$-th entry. The elements of vector ${\bf x}[a,b]$ are comprised of the elements of vector ${\bf x}$, beginning with the $a$-th element and ending at the $b$-th element. $||\bf x||$ denotes the L2-norm of $\bf x$.
For a complex matrix $\bf X$, $\mathbf{X}_{[i,j]}$ represents the element in the $i$-th row and $j$-th column. $\operatorname{Tr}(\bf X)$ and $||\mathbf X||_{F}$ denote its trace and the Frobenius norm of $\bf X$, respectively. 
${\bf x}^{T}$, ${\bf x}^{*}$, ${\bf x}^{H}$, ${\bf X}^{T}$, ${\bf X}^{*}$ and ${\bf X}^{H}$ stand for the transpose, conjugate, and conjugate transpose of vector $\bf x$ and matrix $\bf X$, respectively. 

\section{System Model and Problem Formulation} \label{sec: system model}
\subsection{System Model}

\begin{figure}[t]
 \centering
 \includegraphics[width=0.4\textwidth]{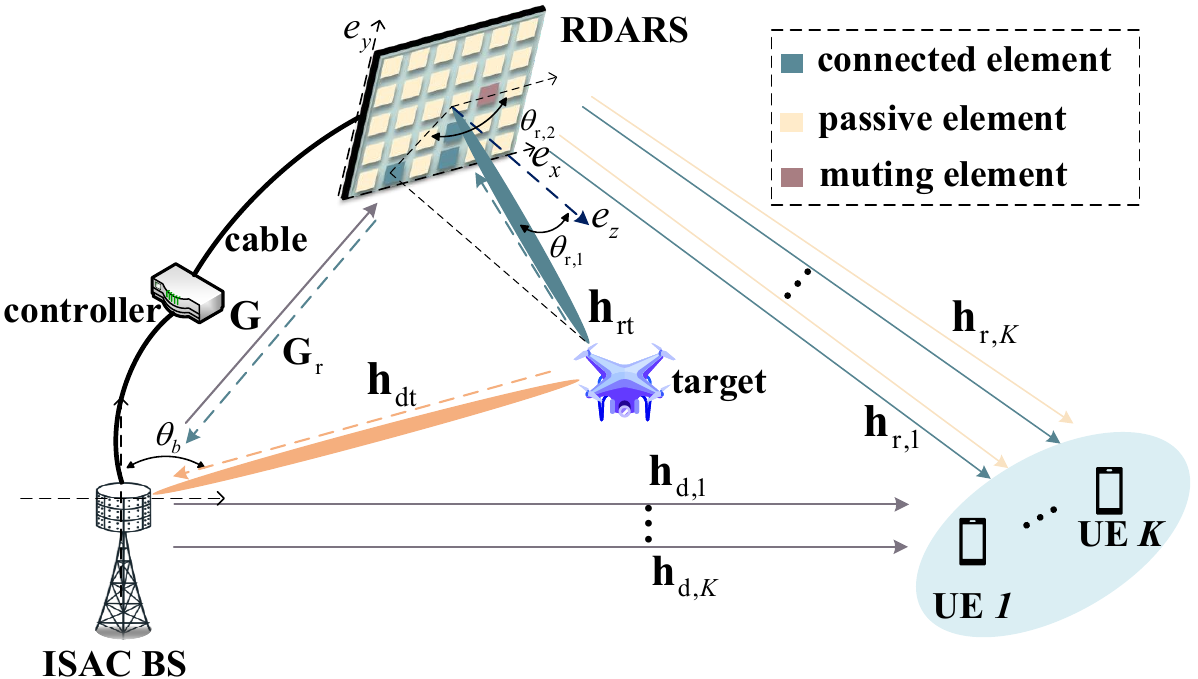}
 \caption{Illustration of the RDARS-aided ISAC system.}
 \label{fig: system architecture}
 \end{figure}
 
As shown in Fig. \ref{fig: system architecture}, the ISAC BS is equipped with a uniform linear array (ULA) comprising $N_{\rm{t}}$ transmit antennas and $N_{\rm{r}}$ receive antennas, and the RDARS adopts a uniform planar array (UPA) architecture with $N$ elements. The transmit and receive arrays are assumed to be sufficiently isolated, such that the self-interference at the ISAC BS transceiver is negligible \cite{keskin2023monostatic}.
Each RDARS element operates in three working modes, including the connection mode, reflection mode, and muting mode. Let the diagonal matrices $\mathbf{A}\in \mathbb{Z}^{N \times N}$ and $\mathbf{B}\in \mathbb{Z}^{N \times N}$ denote the connection and muting mode switching matrices, respectively. 
For simplicity, the muting elements are assumed to dissipate the incident signals through dedicated circuits integrated behind the elements, without reflection or further signal processing.
Specifically, when $\mathbf{A}_{[i,i]} = 1$ and $\mathbf{B}_{[i,i]} = 0$, the $i$-th element works in the connection mode. When $\mathbf{A}_{[i, i]} = 0$ and $\mathbf{B}_{[i, i]} = 1$, the element completely absorbs the signal energy and acts as a muting element.
When $\mathbf{A}_{[i, i]} = 0$ and $\mathbf{B}_{[i, i]} = 0$, the element acts as a passive element that can reflect the signal. 
Let $a$ and $b$ denote the numbers of connected and muting elements, respectively. Accordingly, the number of passive elements is $N_{\rm p} = N-a-b$.
Moreover, the number of single-antenna user equipments (UEs) is $K$, and there exists a single aerial target, which is detected by the BS and RDARS. Let ${\mathbf{X}}_{\rm{b}}\in \mathbb{C}^{N_{\rm t} \times L}$ and ${\mathbf{X}}_{\rm{r}}\in \mathbb{C}^{a \times L}$ denote the DF waveform at ISAC-BS and RDARS, respectively. 
The transmit DF waveform is
$ {\mathbf{X}} =  [{\mathbf{X}}^T_{\rm{b}}, {\mathbf{X}}^T_{\rm{r}}]^T$.
The received signal at the UEs during $L$ time slots can be expressed as
\begin{align}\label{equ: receiving signal Y}
{\bf{Y}} \!\!&=\!\! {\bf{H}}_{\rm d}^H{{\bf{X}}_{\rm b}}\!\! +\!\! {\bf{H}}_{\rm r}^H({\bf{I}}_N \!\!-\!\! {\bf{A}} \!\!- \!\!{\bf{B}}){\bf{\Phi G}}{{\bf{X}}_{\rm b}}\!\! +\!\! {\bf{H}}_{\rm r}^H ({\bf{I}}_N \!\!-\!\! {\bf{B}}) \widetilde {\bf{A}}{{\bf{X}}_{\rm r}} \!\!+\!\! {\bf{W}}\nonumber\\
 &= \!\!{\bf{HX}} + {\bf{W}},
\end{align}
where $\mathbf{G} \in \mathbb{C}^{N \times N_{\rm t}}$ denotes the BS-RDARS channel, ${\mathbf{H}}_{\rm r} =[{{\mathbf{h}}_{{\rm r},1}},{{\mathbf{h}}_{{\rm r},2}}, \cdots ,{{\mathbf{h}}_{{\rm r},K}}] \in {\mathbb{C}^{N \times K}}$ denotes the RDARS-UE channel, ${\mathbf{H}}_{\rm d} \in \mathbb{C}^{N_{\mathrm{t}} \times K}$ denotes the ISAC BS-UE channel, and ${\mathbf{H}}  = [{\mathbf{H}}_{\rm d}^H + {\mathbf{H}}_{\rm r}^H( {{\mathbf{I}}_N - {\mathbf{A}} - {\mathbf{B}}} ){\mathbf{\Phi G}},{\mathbf{H}}_{\rm r}^H( {{\mathbf{I}_N} - {\mathbf{B}}} )\widetilde {\mathbf{A}}] \in \mathbb{C}^{K \times(N_{\mathrm{t}}+a)  }$ denotes the equivalent channel.

Let ${\bf{W}} \in {\mathbb{C}^{K \times L}}$ denote the noise matrix.
By denoting $\mathbf{S} \in {\mathbb{C}^{K \times L}}$ as the desired communication symbol, we can reformulate \eqref{equ: receiving signal Y} as
${\bf{Y}}  = \mathbf{S} + (\mathbf{HX}-\mathbf{S}) + {\bf{W}}$,
where $\mathbf{HX}-\mathbf{S}$ denotes the MUI \cite{liufan_waveform, zhang2025joint_deep_unfolding} and $|\mathbf{S}_{[k,l]}| = 1, \forall k, l$. Therefore, the SINR for the $k$-th user is
$\gamma_{k} = \frac{\mathbb{E}(|{\bf{S}}_{[k,l]}|^2)}{\mathbb{E}(|{\bf{h}}_k {\bf{x}}_l - {\bf{S}}_{[k,l]}|^2) + \sigma ^2}, \forall k$,
where $\mathbb{E}(\cdot)$ denotes the ensemble average with respect to the time slot, and $\sigma ^2$ denotes the noise power.
Due to the normalized entry of $\mathbf{S}$, the minimization of MUI energy is equivalent to the maximization of the achievable rate.

In addition, the DF waveform is utilized for target sensing. Specifically, the DF signal arrives at the target through three distinct paths: 1) a direct link from the ISAC BS to the target; 2) a path from the ISAC BS to the target via reflections from the passive elements of the RDARS; and 3) a path from the RDARS connected elements to the target. 
Then, the BS antennas receive the echo signals from two paths: a direct link from the target to the ISAC BS and a path from the target to the ISAC BS via the RDARS passive elements. 
The steering vector function is defined as
\begin{equation}
    {\mathbf{\hat a}}(N, \tilde{\theta}) = [e^{j\frac{2\pi}{{\lambda}}d\tilde{\theta}\cdot0},e^{j\frac{2\pi}{{\lambda}}d\tilde{\theta}\cdot1}, \cdots, e^{j\frac{2\pi}{{\lambda}}d\tilde{\theta}\cdot(N-1)}]^{T},
\end{equation}
where $\tilde{\theta}$, ${{\lambda}}$, and $d$ denote the spatial frequency, the wavelength, and the antenna spacing, respectively.
Therefore, the received echo signal can be expressed as 
{
\small
\begin{align}
    {{\mathbf{Y}}_{\rm s}} 
    \!\!& =\!\! \sqrt {{\kappa _{\rm{RCS}}}} ( {{\mathbf{h}}_{\rm dt}^* + {\mathbf{G}}_{\rm r}^T({\mathbf{I}_N} - {\mathbf{A}} - {\mathbf{B}}){{\mathbf{\Phi}}^T}{\mathbf{h}}_{\rm rt}^*} ) \nonumber\\
    &[{({{\mathbf{h}}_{\rm dt}^H + {\mathbf{h}}_{\rm rt}^H({\mathbf{I}_N}\!\! -\!\! {\mathbf{A}} \!\!-\!\! {\mathbf{B}}){\mathbf{\Phi G}}}){{\mathbf{X}}_{\rm b}} + {\mathbf{h}}_{\rm rt}^H( {{\mathbf{I}_N}\!\! -\!\! {\mathbf{B}}} )\widetilde {\mathbf{A}}{{\mathbf{X}}_{\rm r}}}]\!\! +\!\! {{\mathbf{W}}_{\rm s}}\nonumber
    \\
    & = \widetilde{\mathbf{H}}{\mathbf{X}} + \mathbf{W}_{\rm s},
\end{align}}where ${\mathbf{h}}_{\rm dt} \in {\mathbb{C}^{{N_{\rm t}} \times 1}}$, ${\mathbf{h}}_{\rm rt}  \in {\mathbb{C}^{{N} \times 1}}$, and ${\mathbf{G}}_{\rm r} \in {\mathbb{C}^{{N} \times {N_{\rm t}}}}$ denote the channels of the ISAC BS-target link, RDARS-target link, and ISAC BS-RDARS link, respectively, and ${\kappa_{\rm{RCS}}}$ denotes the radar cross section (RCS).
Specifically, we have 
${\mathbf{h}}_{\mathrm{dt}} = \kappa_{\mathrm{dt}}\,{\mathbf{\hat a}}(N_{\rm{t}}, \cos(\theta_{\rm b}))$,
${\mathbf{h}}_{\mathrm{rt}} = \kappa_{\mathrm{rt}}\,{\mathbf{\hat a}}_{\mathrm{rt}}\!(N, \vartheta_{{\rm r},1}, \vartheta_{{\rm r},2})=\kappa_{\mathrm{rt}} {\mathbf{\hat a}}(N_{\rm{y}}, \vartheta_{{\rm r},1}) \otimes {\mathbf{\hat a}}(N_{\rm{x}}, \vartheta_{{\rm r},2})$ with $\vartheta_{{\rm r},1} = \cos(\theta_{{\rm r},1})$ and $\vartheta_{{\rm r},2} = \sin(\theta_{{\rm r},2})\sin(\theta_{{\rm r},1})$. Let $N_{\rm y}$ and $N_{\rm x}$ denote the number of RDARS elements along $e_{\rm y}$ and $e_{\rm x}$, respectively, where $e_{\rm y}$ and $e_{\rm x}$ are the unit vectors along the $y$- and $x$-axes of the local coordinate system.
The path-loss coefficients are given by 
$\kappa_{\mathrm{dt}} = \sqrt{\frac{c_0}{d_{\mathrm{bt}}^{\alpha_{\mathrm{bt}}}}}$ and $\kappa_{\mathrm{rt}} = \sqrt{\frac{c_0}{d_{\mathrm{rt}}^{\alpha_{\mathrm{rt}}}}}$, respectively.
In addition, $\theta_{b}$ denotes the angle of arrival (AoA) and $\theta_{{\rm r},1}$ and $\theta_{{\rm r},2}$ represent the azimuth and elevation angles of departure (AoDs), respectively. 
Moreover, we have ${\mathbf{G}}_{\rm r} = {\mathbf{G}}$ when $N_{\mathrm{r}} = N_{\mathrm{t}}$ \cite{song2025fully}.

Then, the received echo signal at the time slot $\ell$ is 
${{\mathbf{y}}_{{\rm s},\ell}} = \widetilde {\mathbf{H}}{{\mathbf{x}}_\ell} + {{\mathbf{w}}_{{\rm s},\ell}}$,
where ${{\mathbf{w}}_{{\rm s},\ell}} \sim {\cal C}{\cal N}( {{\bf{0}},\sigma_{{\rm s}}^2{{\bf{I}}_{N_{\rm{r}}}}} )$ denotes the additive white Gaussian noise (AWGN) with power $\sigma_{s}^2$. Let ${\bf{u}}$ denote the receive beamforming vector and ${\bf W}_{\rm s}=[{{\mathbf{w}}_{{\rm s},1}}, \cdots, {{\mathbf{w}}_{{\rm s},L}}]^T $. 

The sensing SNR at the time slot $\ell$ is given by${\gamma _{{\rm s},\ell}} = \frac{{{{\mathbf{u}}^H}\widetilde {\mathbf{H}}\mathbb{E}\{ {{{\mathbf{x}}_\ell}{\mathbf{x}}_\ell^H} \}{{\widetilde {\mathbf{H}}}^H}{\mathbf{u}}}}{{{{\mathbf{u}}^H}{\mathbf{u}}\sigma _{\rm s}^2}}, 
\forall \ell$.
Given the average covariance matrix ${\mathbf{\bar R}_x} = \frac{1}{L}\sum\nolimits_{l = 1}^L {\mathbb{E}\{ {{{\mathbf{x}}_l}{\mathbf{x}}_l^H}\}}$, the sensing SNR during the $L$ time slots is 
{\small
\begin{equation} \label{equ: ave sensing SNR}
{\overline \gamma_{\rm s}} = \frac{{{{\mathbf{u}}^H}\widetilde {\mathbf{H}}{{{\mathbf{\bar R}}}_x}{{\widetilde {\mathbf{H}}}^H}{\mathbf{u}}}}{{{{\mathbf{u}}^H}{\mathbf{u}}\sigma _{\rm s}^2}}.
\end{equation}}It can be seen that equation \eqref{equ: ave sensing SNR} is a Rayleigh quotient with respect to ${\bf{u}}$. Therefore, the optimal receiving vector ${\bf{u}}^{\rm opt}$ is the eigenvector corresponding to the largest eigenvalue of $\widetilde {\mathbf{H}}{{{\mathbf{\bar R}}}_x}{{\widetilde {\mathbf{H}}}^H}$.

\subsection{Problem Formulation}
For sensing tasks, the frequency-modulated continuous-wave (FMCW) waveform is widely adopted due to its attractive properties, including simple self-interference mitigation, long sensing range, and strong robustness to Doppler shifts, enabling efficient estimation of the range and velocity of multiple targets with low-complexity analog processing \cite{FMCW_OFDM_wang, FMCW_survey_wei, FMCW_LAE_zeng}.
Therefore, the FMCW waveform is chosen as a reference waveform.
In addition, the sensing performance associated with the DF waveform is evaluated by the waveform discrepancy as
$||{\bf{X}} - {\bf{X}}_0||^2_{F}$,
where ${\bf{X}}_0$ denotes the reference FMCW waveform given by
\begin{equation}
\small
\begin{aligned}
    {\bf{X}}_0= \sqrt{\frac{P_{\rm{tot}}}{(N_{\rm{t}} + a)L} }\begin{bmatrix}
{\bf{c}}(\theta_{\rm b}) \\
{\bf{c}}(\theta_{\rm r}) 
\end{bmatrix} \tilde{\bf{x}}^H.
\end{aligned}
\end{equation}
Here, the total transmit power and the steering vectors towards the target direction at the ISAC BS and RDARS are denoted by $P_{\mathrm{tot}}$, 
${\mathbf{c}}(\theta_{\rm b}) = {\mathbf{\hat a}}(N, \cos(\theta_{\rm b}))$, and ${\mathbf{c}}(\theta_{\rm r}) = {\mathbf{E}}(\operatorname{diag}(\bar{\mathbf{A}}) \odot {\mathbf{\hat a}}_{\mathrm{rt}}(N_{\rm{c}}, \vartheta_{{\rm r},1}, \vartheta_{{\rm r},2}))$, respectively, where $\bar{\mathbf{A}} \in \mathbb{R}^{N_{\rm{c}} \times N_{\rm{c}}}$ is obtained based on the connected block and $N_{\rm{c}}$ denotes the total number of RDARS elements in the connected block \cite{ji2025reconfigurable}. Specifically, we define $\mathbf{E}_{[i,j]} = 1$, $i\in \{1, 2, \cdots, a\}$, $j\in \{1, 2, \cdots, N_{\rm{c}}\}$, if $j$ corresponds to the index of the $i$-th nonzero element in $\operatorname{diag}(\bar{\mathbf{A}}) \odot {\mathbf{\hat a}}_{\mathrm{rt}}\!(N_{\rm{c}}, \vartheta_{{\rm r},1}, \vartheta_{{\rm r},2})$, and $\mathbf{E}_{[i,j]} = 0$ otherwise. Let $a$ denote the number of RDARS connected elements.
For simplicity, a rectangular block with its reference point located at the origin is regarded as a connected block. 

Moreover, we have
$\tilde{\mathbf{x}} = [\tilde{x}_{1}, \ldots, \tilde{x}_{L}]^{T}$, where the $l$-th element is expressed as
$ \tilde{x}_{l} = e^{ j2\pi(\frac{f_0}{f_{\rm{s}}}l + \frac{\upsilon}{f^{2}_{\rm{s}}}l^{2})}$
with $f_0$ denoting the carrier frequency, $f_{\rm{s}}$ the sampling frequency, and $\upsilon$ the chirp rate.

Therefore, the joint minimization problem of MUI energy and waveform discrepancy is formulated as
{\small
\begin{subequations}\label{pro: MUI and waveform discrepancy}
 \begin{align}
 \mathop {\min }\limits_{\substack{{{\bf{X}}},{\bf{\Phi }},{\bf{A}}, {\tilde{\bf{A}}},{\bf{B}} }} 
 & \;\; \rho|| \mathbf{HX}-\mathbf{S} ||^{2}_{F} + (1-\rho)||{\bf{X}} - {\bf{X}}_0||^2_{F}\label{of}
 \\
 \;\;\;\textrm{s.t.}\;\;\;\;
 & \operatorname{Tr}({{\bf{X}}}{\bf{X}}^H) \le {P_{\rm{tot}}} ,\label{con: X}\\
 & |\mathbf{\Phi}_{[i,i]}| = 1,  \forall i\in{\mathcal{N}},\label{con: Phi} \\
 &\sum\nolimits_{i = 1}^{N}\!\mathbf{A}_{[i,i]} \!\!= a, \mathbf{A}_{[i,i]} \!\in\! \{0,1\}, \forall i\in{\mathcal{N}}, \label{con: A}\\
 &\sum\nolimits_{i = 1}^{N}\!\tilde{\mathbf{A}}_{[i,l]} \!\!= 1, \tilde{\mathbf{A}}_{[i,l]} \!\!\in\!\! \{0,1\} , \forall l\in{\mathcal{A}}, i\in{\mathcal{N}},\label{con: A tilde} \\
 &  {\bf{A} }= \tilde{{\bf{A}}}\tilde{{\bf{A}}}^{H},\label{con: A +A tilde}\\
  & \sum\nolimits_{i = 1}^{N}\!\mathbf{B}_{[i,i]} \!\!= b, \mathbf{B}_{[i,i]} \!\in\! \{0,1\}, \forall i\in{\mathcal{N}}, \label{con: B}\\
  & \mathbf{A}_{[i,i]}\mathbf{B}_{[i,i]} \neq 1, \forall i\in{\mathcal{N}}\label{con: A*B},
 \end{align}
\end{subequations}}where $\rho$ denotes the weighting factor.
It is observed that problem \eqref{pro: MUI and waveform discrepancy} is non-convex due to the highly coupled variables and unit-modulus constraints. Furthermore, the binary mode switching constraints \eqref{con: A}-\eqref{con: B} further complicate the problem. In addition, constraint \eqref{con: A*B} introduces the extra coupling between the mode switching matrices.

{\section{Special Cases of RDARS-Aided ISAC Systems}\label{sec: Special Cases}}
{To investigate the impact of muting effects and draw useful insights, we consider the special cases of single-UE communication, single-target sensing, and two-UE communication, respectively, by temporarily assuming the basic line-of-sight (LoS) propagation.}
{\subsection{Single-UE Communication}}
For the single-UE communication case, the effective channel is 
$\mathbf h_{\rm eff}^{H}
=[\mathbf h_{\rm d}^H+
\mathbf h_{\rm r}^H(\mathbf I_N\!-\!\mathbf A\!-\!\mathbf B){\bm{\Phi}}\mathbf G,\mathbf h_{\rm r}^H(\mathbf I_N-\mathbf B)\widetilde{\mathbf A}]$. 
With given $\mathbf A$, $\mathbf B$, and $\bm{\Phi}$, the maximum communication SNR is obtained by the applying MRT beamforming, i.e.,
$\mathbf w^{\rm opt}=\sqrt{P_{\rm tot}}
\frac{\mathbf h_{\rm eff}}{\|\mathbf h_{\rm eff}\|_2}$, and we have $\gamma_{\rm c}^{\rm opt} =\frac{P_{\rm tot}}{\sigma^2}
\mathbf h_{\rm eff}^{H}\mathbf h_{\rm eff}$. 
Let $\mathbf M=\mathbf I_N-\mathbf A-\mathbf B$, the channel gain is $\mathbf h_{\rm eff}^{H}\mathbf h_{\rm eff}
=\|\mathbf h_{\rm d}\|_2^2+2\Re\{\mathbf h_{\rm r}^H\mathbf M\bm{\Phi}\mathbf G\mathbf h_{\rm d}\} + \mathbf h_{\rm r}^H\mathbf M\bm{\Phi}\mathbf G\mathbf G^H \bm{\Phi}^H\mathbf M\mathbf h_{\rm r}
+\mathbf h_{\rm r}^H\mathbf A\mathbf h_{\rm r}$.
Under the LoS channel, we have
$\mathbf h_{\rm d}=\kappa_{\rm d}\mathbf a_{\rm d}$,
$\mathbf h_{\rm r}=\kappa_{\rm ru}\mathbf a_{\rm ru}$, and
$\mathbf G=\kappa_{\rm br}\mathbf a_{\rm r}\mathbf a_{\rm t}^H$,
where $\kappa_{\rm d}$, $\kappa_{\rm ru}$, and $\kappa_{\rm br}$ denote the channel coefficients of the BS-UE, RDARS-UE, and BS-RDARS links, respectively. Let
$S_{\Phi}=\mathbf a_{\rm ru}^H(\mathbf I_N-\mathbf A-\mathbf B)\bm{\Phi}\mathbf a_{\rm r}$, the maximum SNR can be expressed as
\begin{equation}\label{equ: optimal SNR single UE}
\small
\begin{aligned} 
\gamma_{\rm c}^{\rm opt}\!\! =\!\! \frac{P_{\rm tot}}{\sigma^2} (
&\!\!\!\!\!\!\underbrace{\kappa^2_{\rm d}N_{\rm t}}_{\textrm{direct-link gain}}\!\!\!\!\!\!\! + \underbrace{2\Re\{\kappa_{\rm ru}^*\kappa_{\rm br}\kappa_{\rm d} S_{\Phi} \mathbf a_{\rm t}^H\mathbf a_{\rm d} \}}_{\textrm{cross gain}}\!\!+ \!\!\underbrace{\kappa^2_{\rm ru}\kappa^2_{\rm br}N_{\rm t} |S_{\Phi}|^2}_{\textrm{reflection gain}}\!\!\!\!
+\!\!\!\!\! \!\!\underbrace{a\kappa^2_{\rm ru}}_{\textrm{distributed gain}}\!\!\!\!\!\!\!).
\end{aligned}
\end{equation}

By applying the optimal passive beamforming, the reflected signals can be coherently combined, and we have $|S_{\Phi}^{\rm opt}|=N-a-b$.
Moreover, the phase of $S_{\Phi}^{\rm opt}$ can be aligned with that of
$\kappa_{\rm ru}^*\kappa_{\rm br}\kappa_{\rm d}\mathbf a_{\rm t}^H\mathbf a_{\rm d}$, where $\bm{\Phi}_{[n,n]} = e^{j(\operatorname{arg}({\bf{a}}_{{\rm ru},n}) -\operatorname{arg}({\bf{a}}_{{\rm r},n}))}$. 
{Thus, the SNR in \eqref{equ: optimal SNR single UE} is independent of the positions of connected elements. 
It is observed that the positions of connected elements can be arbitrarily selected \cite{Wang_RDARS}. 
In the following, we focus on the impacts of muting elements on the resulting SNR.}
Accordingly, the resulting SNR is
\begin{align}\label{equ: maximum SNR for one UE}
\gamma_{\rm c}^{\rm opt}(b)
&=\frac{P_{\rm tot}}{\sigma^2}
(\kappa^2_{\rm d}N_{\rm t}
+\kappa^2_{\rm ru}\kappa^2_{\rm br}N_{\rm t}(N-a-b)^2 \nonumber\\
&+2|\kappa_{\rm ru}^*\kappa_{\rm br}\kappa_{\rm d}
\mathbf a_{\rm t}^H\mathbf a_{\rm d}
|(N-a-b)+a\kappa^2_{\rm ru}),
\end{align}
where $0\le b\le N-a$.

By letting the derivative of $\gamma_{\rm c}^{\rm opt}(b)$ with respect to $b$ be zero, we have
$\frac{d\gamma_{\rm c}^{\rm opt}(b)}{db}
= -\frac{P_{\rm tot}}{\sigma^2}
(2|\kappa_{\rm ru}|^2|\kappa_{\rm br}|^2N_{\rm t}(N-a-b)+2|\kappa_{\rm ru}^*\kappa_{\rm br}\kappa_{\rm d}\mathbf a_{\rm t}^H\mathbf a_{\rm d}|) \le 0$.

Thus, in the single-UE LoS scenario, the maximum communication SNR monotonically decreases with the number of muting elements, which indicates the optimal number for maximizing the SNR is $b^{\rm opt}=0$. This is expected since the increasing number of muting elements will reduce the reflection and cross gains between the direct and reflected link shown in \eqref{equ: maximum SNR for one UE}. 
{\subsection{Single-Target Sensing}}
For a sensing-only scenario with a single target, the target echo channel can be expressed as
$\widetilde{\mathbf H}
=\sqrt{\kappa_{\rm RCS}}\,
\mathbf g_{\rm rx}\mathbf g_{\rm tx}^{H}$, 
where
$\mathbf g_{\rm rx}
=
\mathbf h_{dt}^{*}
+
\mathbf G^{T}
(\mathbf I_N-\mathbf A-\mathbf B)
\bm\Phi^{T}
\mathbf h_{\rm rt}^{*}$,
and
$\mathbf g_{\rm tx}^{H}=[
\mathbf h_{\rm dt}^{H}
+\mathbf h_{\rm rt}^{H}
(\mathbf I_N-\mathbf A-\mathbf B)
\bm\Phi\mathbf G,\;
\mathbf h_{\rm rt}^{H}(\mathbf I_N-\mathbf B)\widetilde{\mathbf A}
]$.

For given transmit covariance matrix $\bar{\mathbf R}_x$, the sensing SNR is given by
$\gamma_{\rm s}
=\frac{
\mathbf u^H\widetilde{\mathbf H}\bar{\mathbf R}_x
\widetilde{\mathbf H}^H\mathbf u
}{\mathbf u^H\mathbf u\sigma_{\rm s}^2
}$.
Since $\widetilde{\mathbf H}$ is a rank-one matrix, the optimal receive beamformer is $\mathbf u^{\rm opt} =  \frac{\mathbf g_{\rm rx}}{||\mathbf g_{\rm rx}||_2}$. Furthermore, with the transmit beam aligned with $\mathbf g_{\rm tx}$, the maximum sensing SNR is $\gamma_{\rm s}^{\rm opt}
=\frac{\kappa_{\rm RCS}P_{\rm tot}}{\sigma^2_{\rm s}}
\|\mathbf g_{\rm rx}\|_2^2
\|\mathbf g_{\rm tx}\|_2^2$.
With 
$\mathbf h_{\rm dt}=\kappa_{\rm dt}\mathbf a_{\rm dt}$,
$\mathbf h_{\rm rt}=\kappa_{\rm rt}\mathbf a_{\rm rt}$, and $\mathbf G=\kappa_{\rm br}\mathbf a_{\rm r}\mathbf a_{t}^H$, and 
let
$\tilde{S}_{\Phi}
= \mathbf a_{\rm rt}^{H}
(\mathbf I_N-\mathbf A-\mathbf B)
{\bm{\Phi}}\mathbf a_{\rm r}$.
Then, we have $\|\mathbf g_{\rm rx}\|_2^2
= \kappa^2_{\rm dt}N_{\rm t}+\kappa^2_{\rm rt}\kappa^2_{\rm br}N_{\rm t}|\tilde{S}_{\Phi}|^2 +
2\Re\{
\kappa_{\rm dt}\kappa_{\rm rt}^{*}\kappa_{\rm br}
\tilde{S}_{\Phi}\mathbf a_{\rm t}^H\mathbf a_{\rm td}
\}$. 
Similarly, we have
$\|\mathbf g_{\rm tx}\|_2^2=|\kappa_{\rm dt}|^2N_{\rm t}+|\kappa_{\rm rt}|^2|\kappa_{\rm br}|^2N_{\rm t}|\tilde{S}_{\Phi}|^2 \nonumber+2\Re\{
\kappa_{\rm dt}\kappa_{\rm rt}^{*}\kappa_{\rm br}\tilde{S}_{\Phi}\mathbf a_{\rm t}^H\mathbf a_{\rm dt}\}+ a|\kappa_{\rm rt}|^2$,
where the direct BS-target and the RDARS-reflected links are coupled through the cross term, i.e., the third term.

By applying the optimal passive beamforming $\bm{\Phi}^{\rm opt}_{[n,n]} = e^{j(\operatorname{arg}({\bf{a}}_{{\rm rt},n}) -\operatorname{arg}({\bf{a}}_{{\rm r},n}))}$, the reflected signals can be coherently combined and the phase of $\tilde{S}_{\Phi}$ can be aligned with that of
$\kappa_{\rm dt}\kappa_{\rm rt}^{*}\kappa_{\rm br}\mathbf a_{\rm t}^H\mathbf a_{dt}$, and thus
$|\tilde{S}_{\Phi}^{\rm opt}|=N-a-b$.

Accordingly, the maximum sensing SNR is
\begin{equation}
\small
\begin{aligned}
\gamma_{\rm s}^{\rm opt}(b)=
\frac{\kappa_{\rm RCS}P_{\rm tot}}{\sigma_{\rm s}^2}
F(b)(F(b)+\underbrace{a\kappa^2_{\rm rt}}_{\textrm{distributed gain}}),
\end{aligned}
\vspace{-10pt}
\end{equation}
with
\begin{equation}
\small
\begin{aligned}
F(b)\!\!=\!\!\!\!\!\!\underbrace{\kappa^2_{\rm dt}N_{\rm t}}_{\textrm{direct-link gain}}\!\!\!\!\!\!+\!\!
\underbrace{\kappa^2_{\rm rt}\kappa^2_{\rm br}N_{\rm t}|\tilde{S}_{\Phi}^{\rm opt}|^2}_{\textrm{reflection gain}} \!\!+\!\!\underbrace{2|
\kappa_{\rm dt}\kappa_{rt}^{*}\kappa_{\rm br}
\mathbf a_{\rm t}^H\mathbf a_{\rm dt}
||\tilde{S}_{\Phi}^{\rm opt}|}_{\textrm{cross gain}}.
\end{aligned}
\end{equation}
Taking the derivative of $F(b)$ with respect to $b$, we have
$\frac{dF(b)}{db}
=
-2\kappa^2_{\rm rt}\kappa^2_{\rm br}N_{\rm t}(N-a-b)
-
2\left|
\kappa_{dt}\kappa_{\rm rt}^{*}\kappa_{\rm br}
\mathbf a_{\rm t}^H\mathbf a_{\rm dt}
\right|$.
Since $0\le b\le N-a$, we have
$\frac{dF(b)}{db}\le 0$.
Furthermore, we have
$\frac{d\gamma_{\rm s}^{\rm opt}(b)}{db}
=\frac{\kappa_{\rm RCS}P_{\rm tot}}{\sigma_{\rm s}^2}\frac{dF(b)}{db}(2F(b)+a|\kappa_{\rm rt}|^2).$
Since $F(b)\ge0$ and $2F(b)+a|\kappa_{\rm rt}|^2>0$, it follows that
$\frac{d\gamma_{\rm s}^{\rm opt}(b)}{db}\le0$.

It is observed that the maximum sensing SNR monotonically decreases with the number of muting elements, and $b^{\rm opt}=0$ is optimal to maximize the sensing SNR.
This is expected since similar to the single-UE communication case, the increase in the number of muting elements reduces the reflection gain and cross gain.

{
It is worth noting that, although increasing the number of muting elements degrades the single-UE and single-target performance, it reshapes the passive beam pattern. To this end, we adopt the mainlobe-to-sidelobe ratio (MSR) to characterize the impact of muting elements on both communication and sensing, where a higher MSR indicates lower sidelobe leakage, leading to reduced MUI and sensing ambiguity \cite{MSR_2026, jiu2014knowledge}.
Specifically, the MSR is defined as $\eta_{\rm MSR} = \frac{P^{\rm peak}_{\rm ML}}{P^{\rm peak}_{\rm SL}}$,
where $P^{\rm peak}_{\rm ML}$ and $P^{\rm peak}_{\rm SL}$ denote the peak powers of mainlobe and sidelobe, respectively.
We define the beam power of passive elements as $P = |{\bf a}^H(\bar \theta) ({\bf I}_N - {\bf A} - {\bf B}) {\bm \varphi}|$, where ${\bf a}^H(\bar \theta)$ denotes the effective steering vector with $\bar \theta$ and ${\bm \varphi}$ being the spatial frequency and passive beam. Therefore, we have $P \le N - a - b$. For ease of presentation, the width of mainlobe is defined as the null-to-null beam width, i.e., $\frac{2}{N-a-b}$.}

\begin{figure}[htbp]
\centering
\subfloat[$b=0$.]{\includegraphics[width=0.49\columnwidth]{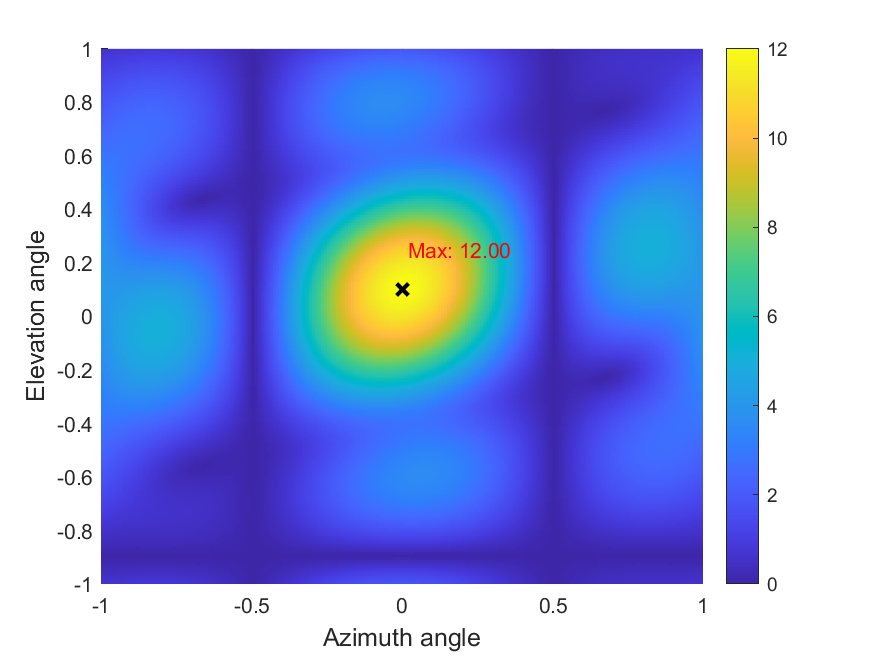}}
\hfill
\subfloat[$b=2$.]{\includegraphics[width=0.49\columnwidth]{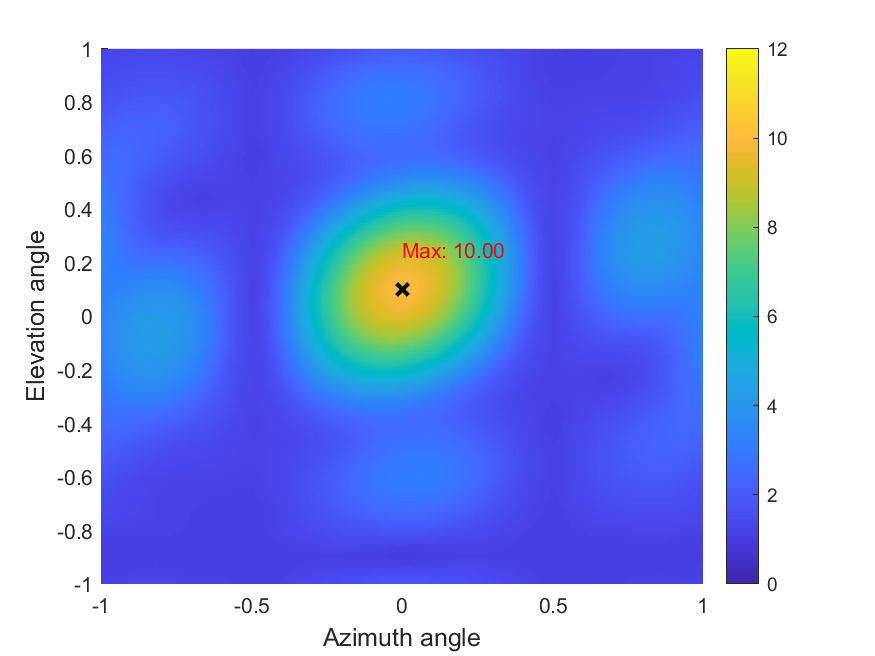}}
\hfill
\subfloat[$b=4$.]{\includegraphics[width=0.5\columnwidth]{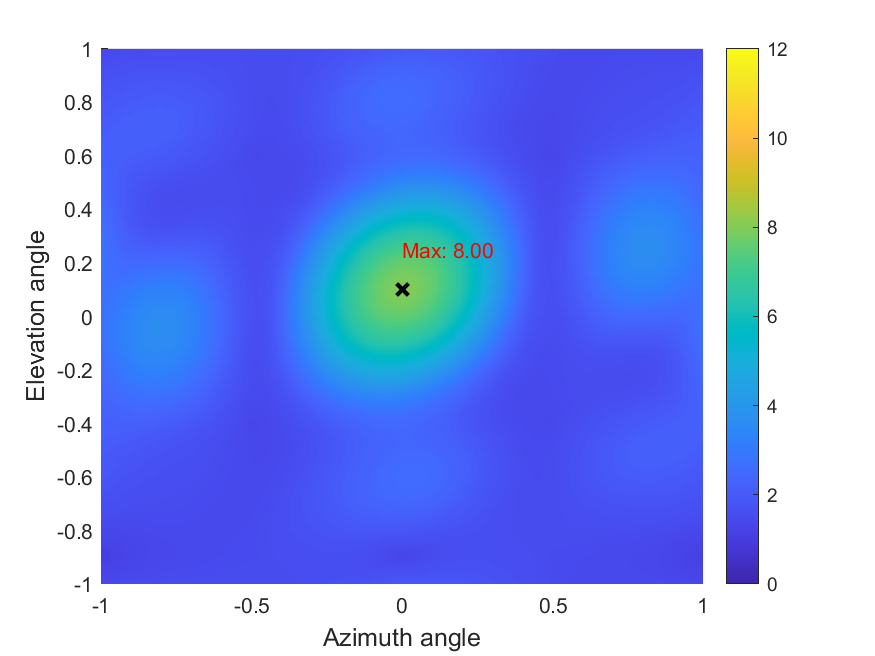}}
\hfill
\subfloat[$b=8$.]{\includegraphics[width=0.5\columnwidth]{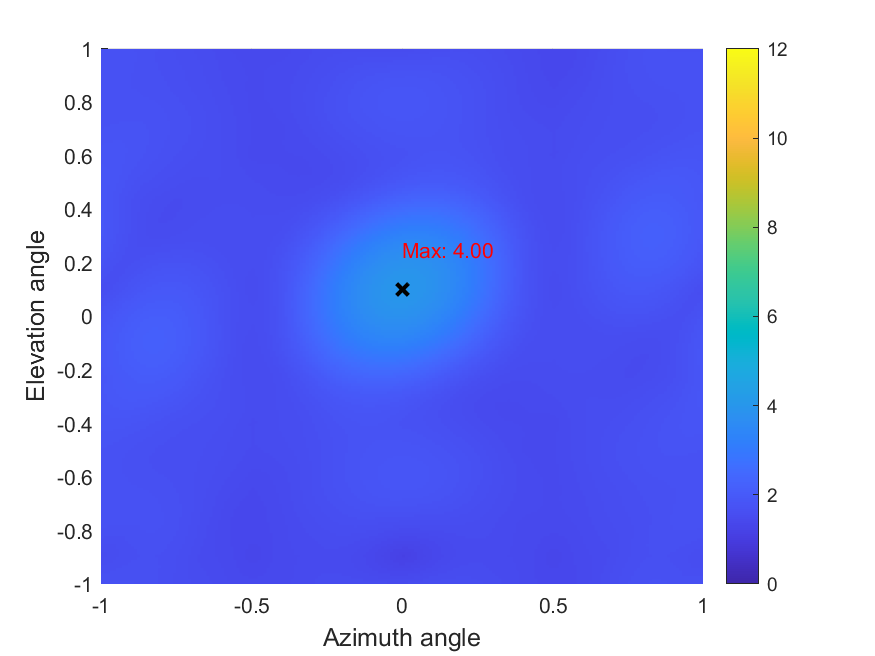}}
\caption{{Comparison of beam pattern gains for passive elements under different numbers of muting elements $b$. The RDARS is a $4\times 4$ UPA with $a=4$ connected elements.}}
\label{fig: UPA beam pattern}
\vspace{-10pt}
\end{figure}

\begin{figure}[htbp]
\centering
\subfloat[Beam pattern versus channel angles.]
{\includegraphics[width=0.5\columnwidth]{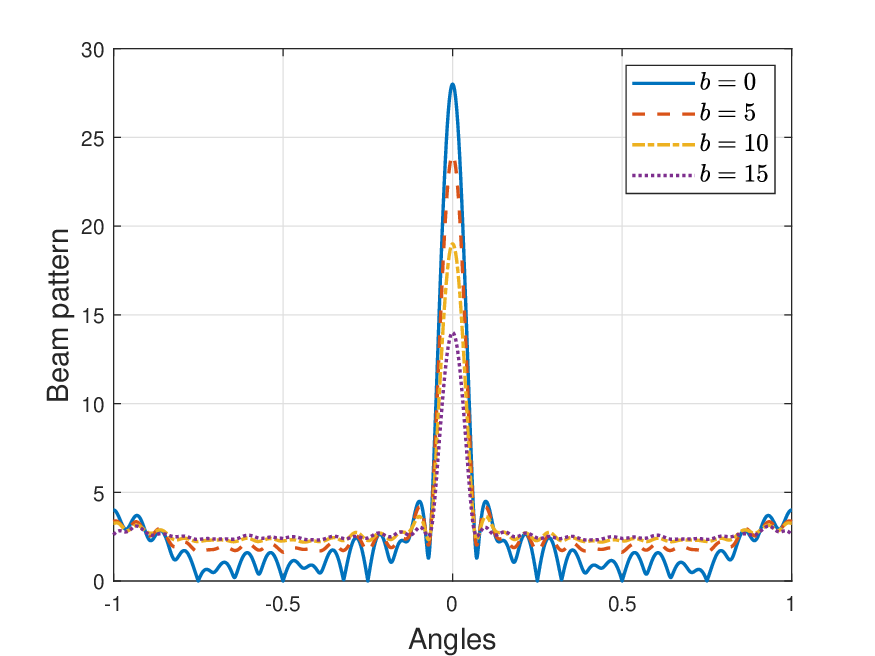}\label{fig: ULA beam pattern a}}
\hfill
\subfloat[MSR versus the number of muting elements.]{\includegraphics[width=0.5\columnwidth]{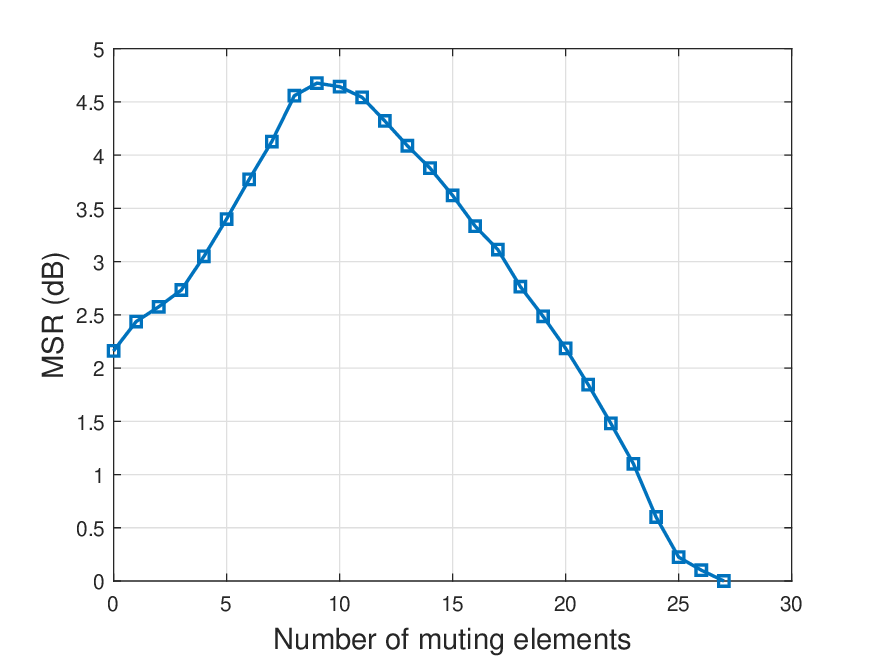}}
\caption{{The RDARS is a 32-element ULA with $a=4$ connected elements. The beam pattern for passive elements is shown under different numbers of muting elements $b$.}}
\label{fig: ULA beam pattern}
\vspace{-15pt}
\end{figure}

{For the single-UE/target case, the beam patterns of passive elements as examples are shown in Figs. \ref{fig: UPA beam pattern} and \ref{fig: ULA beam pattern}. It is observed that the mainlobe strength is significantly reduced and the sidelobe gain increases when the number of muting elements increases, as expected. However, as $b$ increases, the mainlobe becomes wider while the sidelobes become flatter, as shown in Figs. \ref{fig: UPA beam pattern} and \ref{fig: ULA beam pattern}\subref{fig: ULA beam pattern a}. In particular, the MSR first increases and then decreases as $b$ increases, thus existing a trade-off for the MUI mitigation and sensing performance enhancement for multi-UE and general ISAC cases.}

{\subsection{Two-UE Communication}}
For the two-UE communication, the desired signal power of UE $k$ is $\|\mathbf h_k\|_2^2 = E_k+N_{\rm p}^2G_k$, where $E_k=\kappa^2_{{\rm d},k}N_{\rm t}+a\kappa^2_{{\rm r},k}$ denotes the channel gain arising from BS direct link and the RDARS connected elements, and $G_k=\frac{|\kappa_{{\rm r},k}|^2|\kappa_{\rm br}|^2N_{\rm t}}{N^2_{\rm p}}|\mathbf a_{{\rm r},k}^{H}(\mathbf I_N-\mathbf A-\mathbf B)\bm{\Phi}\mathbf a_{\rm r}|^2$
accounts for the effective reflection gain of UE $k$, with $\kappa_{{\rm d},k}$, $\kappa_{{\rm r},k}$, and $\kappa_{\rm br}$ being the path loss coefficients of the BS-UE $k$, RDARS-UE $k$, and BS-RDARS links, respectively. Let $g_{{\rm r}, k} = \frac{|\mathbf a_{{\rm r},k}^{H}(\mathbf I_N-\mathbf A-\mathbf B)\bm{\Phi}\mathbf a_{\rm r}|}{N_{\rm p}}$ denote the normalized passive reflection gain for UE $k$.

By applying the simple MRT beamforming to UE $k$, i.e., $\mathbf w_k=\frac{\mathbf h_k}{\|\mathbf h_k\|_2}$,
the SINR of UE $k$ is $\gamma_k =\frac{P_k\|\mathbf h_k\|_2^2}{\sigma^2+P_j|\mathbf h_k^H\mathbf h_j|^2 /\|\mathbf h_j\|_2^2},\;j, k = 1,2, j\neq k$,
where $P_k$ denotes the transmit power of UE $k$. Due to the shared passive beamforming, the reflection-link interference between the two UEs becomes significant. Therefore, focusing on the interference dominated by the reflection link, we have the following proposition.
\begin{Proposition}\label{Propo: 1}
    When $N_{\rm p}^2\sqrt{C_{k,j}} \gg \sqrt{E_kE_j} + N_{\rm p}\sqrt{E_kG_j} + N_{\rm p}\sqrt{G_kE_j}$ where $C_{{\rm{k}},j}=|\kappa_{{\rm{r}},k}|^2|\kappa_{{\rm{r}},j}|^2|\kappa_{\rm br}|^4N_{\rm t}^2 
    |g_{{\rm r}, k}|^2|g_{{\rm r}, j}|^2$, the interference is dominated by the reflection link. Accordingly, the interference term is approximated as
    {\small
    \begin{align}
    P_j\frac{|\mathbf h_k^H\mathbf h_j|^2}{\|\mathbf h_j\|_2^2} \approx P_j\frac{N^{4}_{\rm p}C_{k,j}}{E_j+N^{2}_{\rm p}G_j},
    \end{align}}where $|\mathbf h_k^H\mathbf h_j|^2 = N^{4}_{\rm p} C_{k,j}$ denotes the reflection link correlation coefficient of two UEs.
\end{Proposition}
{\it{Proof:}} Please refer to Appendix A.
~$\hfill\blacksquare$

With Proposition \ref{Propo: 1}, the SINR of UE $k$ can be approximated as
$\gamma_k =\frac{P_k(E_k+N_{\rm p}^2G_k)}{\sigma^2+P_j(N_{\rm p}^4C_{k,j})/(E_j+N_{\rm p}^2G_j)}$.
Let ${\tilde{N}_{\rm p}}=N_{\rm p}^2$, the SINR can be rewritten as
$\gamma_k ({\tilde{N}_{\rm p}}) =\frac{P_k(E_k+{\tilde{N}_{\rm p}}G_k)}{\sigma^2+
P_j\frac{{\tilde{N}^2_{\rm p}}C_{k,j}}{E_j+{\tilde{N}_{\rm p}}G_j}
}$.

Based on the above analysis, we have the following proposition.
\begin{Proposition}\label{Propo: 2}
    When $\frac{P_j}{\sigma^2}>\tau_{k,j}$, the optimal number of muting elements is 
    {\small
    \begin{align}
        b_k^{\rm opt}=\operatorname{round}(N-a-\sqrt{{\tilde N}^{\rm opt}_{{\rm p},k}}),
    \end{align}}where $\tau_{k,j} = \frac{(E_j+(N-a)^2 G_j)^2}{G_j (N-a)^2 [2E_kE_j+(N-a)^2(E_kG_j+G_kE_j)]}$ denotes the muting gain threshold and ${\tilde N}^{\rm opt}_{{\rm p},k}=\frac{- c_{k,1} - \sqrt{c_{k,1}^2 - 4 c_{k,2} c_{k,0}}}{2 c_{k,2}}$ with  $c_{k,0}=G_k\sigma^2E_j^2$, $c_{k,1}=2G_k\sigma^2E_jG_j-2P_jC_{k,j}E_kE_j$, and $c_{k,2}=G_k\sigma^2G_j^2-P_jC_{k,j}(E_kG_j+G_kE_j)$.
    \end{Proposition}
{\it{Proof:}} Please refer to Appendix B.
~$\hfill\blacksquare$

It is observed that the condition $\frac{P_j}{\sigma^2}>\tau_{k,j}$ can be satisfied. Therefore, the muting elements yield a positive SINR gain.
In addition, Proposition \ref{Propo: 1} reveals the tradeoff between the MUI mitigation and reflection gain introduced by the muting mode. Under the above conditions, the SINR of UE $k$ first increases and then decreases with the number of muting elements. This is because when $b$ is small, a strong reflection link correlation coefficient leads to a severe MUI, and thus increasing $b$ will reduce the reflection link correlation coefficient and the MUI. However, when the MUI is well suppressed for a moderately large $b$, further increasing $b$ will incur reflection gain loss. 
\section{General ISAC Scenario}\label{sec: APDD}
In this section, we propose an APDD algorithm to solve the problem \eqref{pro: MUI and waveform discrepancy} by jointly designing waveform, passive beamforming, and mode switching matrices. In the following, each variable is optimized via the individual block.
\subsection{Waveform Optimization}
First, with fixed $\{ {\bf{\Phi }},{\bf{A}}, {\tilde{\bf{A}}},{\bf{B}} \}$, the MUI energy can be expressed as
$\| { {\mathbf{HX}} - {\mathbf{S}}}\|_F^2 = \sum\nolimits_{k = 1}^K {\| {( {{{\mathbf{I}}_L} \otimes {{\mathbf{h}}_k}} ){\mathbf{x}} - {{\mathbf{s}}_k}} \|_2^2}$,
where ${\mathbf{H}} = {[{\mathbf{h}}_{1}^T,{\mathbf{h}}_{2}^T, \cdots ,{\mathbf{h}}_{K}^T]^T}$ with ${\mathbf{h}}_{k}\in \mathbb{C}^{1\times (N_{\rm t} + a)}$, ${\mathbf{S}} = {[{{\mathbf{s}}_1},{{\mathbf{s}}_2}, \cdots ,{{\mathbf{s}}_K}]^T}$ with ${{\mathbf{s}}_k}\in\mathbb{C}^{L \times 1}$, and ${\bf{x}} = \operatorname{vec}({\mathbf{X}}) \in\mathbb{C}^{L(N_{\rm t} + a) \times 1}$.
Therefore, the original problem is rewritten as
\begin{subequations}\label{pro: X}
 \begin{align}
 \mathop {\min }\limits_{\substack{{{\bf{X}}} }} 
 & \; \rho\sum\nolimits_{k = 1}^K {\|{({{{\mathbf{I}}_L} \!\otimes\! {{\mathbf{h}}_k}}){\mathbf{x}} \!- \!{{\mathbf{s}}_k}}\|_2^2}\! +\! (1\!-\!\rho)||{\bf{x}} - {\bf{x}}_0||^2_{2}
 \\
 \textrm{s.t.}\;
 &\; {\bf{x}}^H {\bf{x}} \le {P_{\rm{tot}}},\label{con: x}
 \end{align}
\end{subequations}
where ${\bf{x}}_0 = \operatorname{vec}({\bf{X}}_0)$.
The scaled augmented Lagrangian function of the objective function in \eqref{pro: X} is
${\mathcal{L}_{ {\mathbf{x}}}}( {{\mathbf{x}},{\tilde{\lambda}} } ) = \rho \sum\nolimits_{k = 1}^K \| {( {{{\mathbf{I}}_L} \otimes {{\mathbf{h}}_k}} ){\mathbf{x}} - {{\mathbf{s}}_k}} \|_2^2  + (1 - \rho )\|{\mathbf{x}} - {{\mathbf{x}}_0} \|_2^2+ {\tilde{\lambda}}( {{{\mathbf{x}}^H}{\mathbf{x}} - {P_{\rm{tot}}}})$,
where ${\tilde{\lambda}}$ represents the Lagrangian multiplier.
Based on the Karush-Kuhn-Tucker (KKT) condition, we have
$\frac{{\partial {\mathcal{L}_{ {\mathbf{x}}}}( {{\mathbf{x}},{\tilde{\lambda}} })}}{{\partial {\mathbf{x}}}} 
   =\! \rho \sum\nolimits_{k = 1}^K {{\mathbf{C}}_k^T( {{\mathbf{C}}_k^ * {{\mathbf{x}}^*} - {\mathbf{s}}_k^*})} +\! (1-\rho)({{\mathbf{x}}^*}\!-\!{\mathbf{x}}_0^*) + {\tilde{\lambda}} {{\mathbf{x}}^*}\! = {\mathbf{0}}$, 
where ${\mathbf{C}}_k ={{\mathbf{I}}_L} \otimes {{\mathbf{h}}_k}$.
\begin{figure*}[!t]
{\footnotesize
{
\vspace{-10pt}
  \begin{align} 
     & {{\mathbf{x}}^{\rm{opt}}} = \frac{{\sqrt {{P_{\rm{tot}}}} {{( {\rho \sum\nolimits_{k = 1}^K {{\mathbf{C}}_k^H{{\mathbf{C}}_k}}  + ( {1 - \rho   } ){{\mathbf{I}}_{L({N_{\rm t}} + a)}}} )}^{-1}}( {\rho \sum\nolimits_{k = 1}^K {{\mathbf{C}}_k^H{{\mathbf{s}}_k}}  + (1 - \rho ){{\mathbf{x}}_0}} )}}{{{{\| {{{( {\rho \sum\nolimits_{k = 1}^K {{\mathbf{C}}_k^H{{\mathbf{C}}_k}} + ({1 -\rho}){{\mathbf{I}}_{L({N_{\rm t}} + a)}}} )}^{-1}}( {\rho \sum\nolimits_{k = 1}^K {{\mathbf{C}}_k^H{{\mathbf{s}}_k}}  + (1 - \rho ){{\mathbf{x}}_0}})}\|}_2}}}. \label{equ: x_opt_P}
\end{align}}}
{\noindent} \rule[-0pt]{18.3cm}{0.05em}
\vspace{-10pt}
\end{figure*}
Without loss of generality, ${\tilde{\lambda}}$ can be set to 0 and ${{\mathbf{x}}^{\rm{opt}}}$ can be normalized to satisfy the constraint \eqref{con: x}, which is given by \eqref{equ: x_opt_P}.
\subsection{Phase Shift Optimization}
With fixed  $\{ {\bf{X}},{\bf{A}}, {\tilde{\bf{A}}},{\bf{B}}\}$, the MUI energy is rewritten as
$\| {{\mathbf{HX}} - {\mathbf{S}}} \|_F^2= \sum\nolimits_{k = 1}^K {\| {( {{{\mathbf{I}}_L} \otimes( {{{\bm{\phi}} ^H}{{\widetilde {\mathbf{H}}}_{{\rm r},k}}} )} ){{\mathbf{x}}_b} + {{\widetilde {\mathbf{s}}}_k}} \|_2^2}$,
where $\widetilde {\mathbf{H}}_{{\rm r},k} = \operatorname{diag}( {{\mathbf{h}}_{{\rm r},k}^H} )( {\mathbf{I}_N} - {\mathbf{A}} - {\mathbf{B}}){\mathbf{G}}$ and $\widetilde {\mathbf{s}}_k^T = {\mathbf{h}}_{d,k}^H{{\mathbf{X}}_b} + {\mathbf{h}}_{{\rm r},k}^H( {\mathbf{I}_N} - {\mathbf{B}})\widetilde {\mathbf{A}}{{\mathbf{X}}_{\rm r}} - {\mathbf{s}}_k^T$.
Therefore, problem \eqref{pro: MUI and waveform discrepancy} can be transformed into
{{\small}
\begin{align}\label{pro: phi}
\mathop {\min }\limits_{\substack{ {\bf{\Phi }}}} 
\; \rho \sum\nolimits_{k = 1}^K {\|{( {{{\mathbf{I}}_L} \otimes ( {{\bm\phi^H}{{\widetilde {\mathbf{H}}}_{{\rm r},k}}} )} ){{\mathbf{x}}_b} + {{\widetilde {\mathbf{s}}}_k}} \|_2^2} \;\; \textrm{s.t.}
\; \eqref{con: Phi}.
 \end{align}}

To solve this problem, a power iteration-based algorithm is applied. 
Specifically, problem \eqref{pro: phi} can be equivalently transformed into
\begin{align}\label{pro: phi PI}
 \mathop {\min }\limits_{\substack{ {\bf{\Phi }} }} 
 \; {{\bm \phi} ^H}{\mathbf{D}}{\bm\phi}  + {\bm \phi ^H}{\bm{\beta}} + {{\bm{\beta}}^H} {\bm\phi} 
\;\;\; \textrm{s.t.}
 \; |\phi_{i}| = 1,  \forall i\in{\mathcal{N}},
 \end{align}
where ${\mathbf{D}} = \rho \sum\nolimits_{k = 1}^K \sum\nolimits_{l = 1}^L {{\mathbf{d}}_{k,l}}{\mathbf{d}}_{k,l}^H $ and ${\bm{\beta}} = \rho \sum\nolimits_{k = 1}^K \sum\nolimits_{l = 1}^L {{\mathbf{d}}_{k,l}}\tilde s_{k,j}^H$.
Let ${\mathbf{p}} = {[{\bm{\phi}} ,q]^T}$, where $p_n$ is the $n$-th element of ${\mathbf{p}}$, and $q$ is an auxiliary variable.
Therefore, problem \eqref{pro: phi PI} can be equivalently formulated as \cite{NT_ISAC, ISAC-NET}
\begin{equation}
\small
\begin{aligned}\label{pro: phi PI D_tilde}
 \mathop {\max }\limits_{\substack{ {\bf{p }} }} 
 \; {{\bf{p}} ^H}\widetilde {\mathbf{D}} {\bf{p }} 
\;\;\textrm{s.t.}
\; | {{p_i}}| = 1,i = 1,...,N + 1,
 \end{aligned}
\end{equation}
where ${\widetilde {\mathbf{D}}} = [ - {\mathbf{D}},-{\bm{\beta }}; -{{\bm{\beta }}^H},0]$. It is observed that the optimization problem \eqref{pro: phi PI D_tilde} is a unimodular quadratic program and can be solved by the power iteration algorithm. Specifically, the value of $\bf{p}$ in the $(t+1)$-th iteration is
\begin{equation}\label{equ: PI p}
    {{\mathbf{p}}^{(t + 1)}} = {e^{j\arg ((\widetilde {\mathbf{D}} + \varepsilon {{\mathbf{I}}_{N + 1}}){{\mathbf{p}}^{(t)}})}},
\end{equation}
where $\varepsilon$ is introduced to guarantee that $\widetilde {\mathbf{D}} + \varepsilon {{\mathbf{I}}_{N + 1}}$ is a positive definite matrix. 
Then, the optimized ${\bm \phi}$ is ${\bm\phi}^{\rm{opt}}  = {e^{j\arg (\frac{{{\mathbf{p}}[1:N]}}{{{p_{N + 1}}}})}}$ when the iteration \eqref{equ: PI p} converges.
\subsection{Connection Mode Switching Matrix Optimization}
With fixed $\{ {\bf{X}}, {\bf{\Phi }},{\widetilde{\bf{A}}},{\bf{B}} \}$, the penalty term $\xi$ is introduced to relax the constraint $\eqref{con: A +A tilde}$. Therefore, by temporarily ignoring constraint \eqref{con: A*B}, problem \eqref{pro: MUI and waveform discrepancy} can be reduced to
 \begin{align}\label{pro: A dual}
 \mathop {\min }\limits_{\substack{ {\mathbf{A}} }} 
 \; {\mathcal{L}_{\mathbf{A}}}\left( {{\mathbf{A}},\xi} \right) 
 \;\;\textrm{s.t.} \;\eqref{con: A},
 \end{align}
where the Lagrangian function with respect to ${\mathbf{A}}$ is given by ${\mathcal{L}_{\mathbf{A}}}( {{\mathbf{A}},\xi }) 
      = \rho \sum\nolimits_{k = 1}^K \| {\mathbf{h}}_{{\rm d},k}^H{{\mathbf{X}}_{\rm b}} + {\phi ^H}\operatorname{diag}( {{\mathbf{h}}_{{\rm r},k}^H} )( {{\mathbf{I}}_N} - {\mathbf{A}} - {\mathbf{B}}){\mathbf{G}}{{\mathbf{X}}_{\rm b}} + {\mathbf{h}}_{{\rm b},k}^H( {\mathbf{I}_N} - {\mathbf{B}} )\widetilde {\mathbf{A}}{{\mathbf{X}}_{\rm b}} - {\mathbf{s}}_k^T \|_2^2
    + \frac{1}{{2\xi}}\|{\mathbf{A}} - \widetilde {\mathbf{A}}{{\widetilde {\mathbf{A}}}^H}\|_F^2$.
By defining ${\bf{a}} = \operatorname{diag}(\bf{A})$, problem \eqref{pro: A dual} can be expressed as
\begin{align}\label{pro: a}
 \mathop {\min }\limits_{\substack{ {\mathbf{a}} }} 
  \; {\mathcal{L}_{\mathbf{a}}}( {{\mathbf{a}},\xi } ) 
\;\;\textrm{s.t.}
 \; a_i \in \{0,1\}, \forall i \in \mathcal{N},
\end{align}
where ${\mathcal{L}_{\mathbf{a}}}( {{\mathbf{a}},\xi } ) = \rho \sum\nolimits_{k = 1}^K {\| { - {{\mathbf{a}}^T}{\mathbf{\Phi }}\operatorname{diag} ( {{\mathbf{h}}_{{\rm r},k}^H} ){\mathbf{G}}{{\mathbf{X}}_b} + {\mathbf{r}}_{{1,k}}^T} \|_2^2}  + \frac{1}{{2\xi }}( {{\mathbf{r}}_2^T{\mathbf{a}} + {r_3}} )$ with 
${\mathbf{r}}_{{1,k}}^T =  {\phi ^H}\operatorname{diag} ( {{\mathbf{h}}_{{\rm r},k}^H} )( {{{\mathbf{I}}_N} - {\mathbf{B}}} ){\mathbf{G}}{{\mathbf{X}}_b} - {\mathbf{h}}_{{\rm r},k}^H( {{\mathbf{I}_N} - {\mathbf{B}}})\widetilde {\mathbf{A}}{{\mathbf{X}}_{\rm r}} - {\mathbf{s}}_k^T +{\mathbf{h}}_{d,k}^H{{\mathbf{X}}_b}$, ${{\mathbf{r}}_2} = {\operatorname{diag}}( {{{\mathbf{I}}_N} - 2\widetilde {\mathbf{A}}{{\widetilde {\mathbf{A}}}^H}} )$, and ${r_3} = {\operatorname{Tr}}( \widetilde{\mathbf{A}}{{\widetilde {\mathbf{A}}}^H} )$, respectively. 
According to the quadratic form ${{\mathbf{a}}^T}\widetilde {\mathbf{E}}{\mathbf{a}}$ with $\widetilde{\mathbf{E}} = \sum\nolimits_{k = 1}^K {{{\mathbf{E}}_k}{\mathbf{E}}_k^H} $ and ${{\mathbf{E}}_k} = {\mathbf{\Phi }}\operatorname{diag} ( {{\mathbf{h}}_{{\rm r},k}^H} ){\mathbf{G}}{{\mathbf{X}}_b}$, the MM technique is applied to obtain a tractable surrogate function. Then, for given the resulting value ${\bf a}^{(t)}$ at the $t$-th step, we have
\begin{equation}
    {{\mathbf{a}}^T}\widetilde {\mathbf{E}}{\mathbf{a}}\!\! \leq \!\!{{\mathbf{a}}^T}{{\mathbf{\Lambda }}_e}{\mathbf{a}}\! +\! 2\Re \{ {{{\mathbf{a}}^T}(\widetilde {\mathbf{E}} \!\!-\!\! {{\mathbf{\Lambda }}_e}){{\mathbf{a}}^{(t)}}} \}\! +\! {{\mathbf{a}}^{(t)}}^T({{\mathbf{\Lambda }}_e}\!- \!\widetilde {\mathbf{E}}){{\mathbf{a}}^{(t)}},
\end{equation}
where ${{\mathbf{\Lambda }}_e} =\bar{\lambda}_{\rm{max}}(\widetilde {\mathbf{E}}){\bf I}_{N}$ and $\bar{\lambda}_{\rm{max}}(\widetilde {\mathbf{E}}) $ denotes the maximum eigenvalue of $\widetilde {\mathbf{E}}$.
Therefore, the Lagrangian function can be equivalently written as 
${\widetilde {\mathcal{L}}_{\mathbf{a}}}( {{\mathbf{a}},\xi }) = - 2\rho \Re \{{\sum\nolimits_{k = 1}^K {{\mathbf{r}}_{^1}^T{\mathbf{E}}_k^H} {\mathbf{a}}}\} + 2\rho \Re \{ {{{\mathbf{a}}^T}( {\widetilde {\mathbf{E}} - {\Lambda _e}} ){{\mathbf{a}}^{(t)}}} \}+ \frac{1}{{2\xi}}( {{\mathbf{r}}_2^T{\mathbf{a}} + {r_3}} ) = \Re\{{{\mathbf{r}}_4^H{\mathbf{a}}}\}$,
where ${{\mathbf{r}}_4} =  - 2\rho \sum\nolimits_{k = 1}^K {{{\mathbf{E}}_k}{\mathbf{r}}_{{1,k}}^*}  + 2\rho( {\widetilde {\mathbf{E}} - {\Lambda _e}}){{\mathbf{a}}^{(t)}} + \frac{1}{{2\xi }}{{\mathbf{r}}_2}$.

Therefore, problem \eqref{pro: a} can be reformulated as
 \begin{align}\label{pro: a r4}
 \mathop {\min }\limits_{\substack{ {\mathbf{a}} }} 
 \; \Re \left\{ {{\mathbf{r}}_4^H{\mathbf{a}}} \right\} 
 \;\;\textrm{s.t.} 
\; a_i \in \{0,1\}, \forall i \in \mathcal{N}.
 \end{align}
Let $\mathcal{M}$ denote the set of indices corresponding to the $a$ smallest entries of $\Re \{{{\mathbf{r}}_4} \}$.

Accordingly, the optimal solution to problem \eqref{pro: a r4} is given by
{\small
\begin{equation} \label{equ: opt_a}
{a}^{\mathrm{opt}}_{i} =
\begin{cases}
 1, &\Re\{{{\bf{r}}_4}\}_{[i]} \in \mathcal{M} , \\
 0, &\Re\{{{\bf{r}}_4}\}_{[i]}  \notin \mathcal{M}. 
\end{cases}
\end{equation}}

It is observed from \eqref{equ: opt_a} that we need to find the first $a$ minimum elements from $\Re \left\{ {{\mathbf{r}}_4} \right\}$, i.e., $a_i = 1$, where the corresponding indices specify the elements working in the connection mode.


\subsection{Equivalent Connection Mode Switching Matrix Optimization}
With ${\mathbf{r}}_{5,k}^T = {\bm{\phi} ^H}\operatorname{diag}( {{\mathbf{h}}_{{\rm r},k}^H} )( {{{\mathbf{I}}_N}\!\!-\!\! {\mathbf{A}} \!\!-\!\! {\mathbf{B}}} ){\mathbf{G}}{{\mathbf{X}}_{\rm b}} +  {\mathbf{h}}_{{\rm d},k}^H{{\mathbf{X}}_{\rm b}}\!-\! {\mathbf{s}}_k^T$ and ${\mathbf{\tilde a}} = \operatorname{vec} ( {\widetilde {\mathbf{A}}} ) = {[ {{{\widetilde {\mathbf{A}}}^T}[:,1], \cdots ,{{\widetilde {\mathbf{A}}}^T}[:,a]}]^T}$, the Lagrangian function of problem \eqref{pro: MUI and waveform discrepancy} with respect to ${\widetilde {\mathbf{A}}}$ is given by
${\mathcal{L}_{\widetilde {\mathbf{A}}}}( {\widetilde {\mathbf{A}},\xi }) 
 = \rho \sum\nolimits_{k = 1}^K( {( {{{\mathbf{X}}_{\rm r}}{\mathbf{r}}_{^{5,k}}^*} ) \otimes( {( {{\mathbf{I}_N} - {\mathbf{B}}} ){\mathbf{h}}_{{\rm r},k}^*})} )^H\widetilde {\mathbf{a}} +{{\widetilde {\mathbf{a}}}^H}(( {{{\mathbf{X}}_{\rm r}}{\mathbf{r}}_{^{5,k}}^*} )\otimes(( {\mathbf{I}_N}  -  {\mathbf{B}} ){\mathbf{h}}_{{\rm r},k}^*)) 
 +{{\widetilde {\mathbf{a}}}^H}( ( {{\mathbf{X}}_{\rm r}}{\mathbf{X}}_{\rm r}^H )\otimes( ( {\mathbf{I}_N}  - {\mathbf{B}} ){\mathbf{h}}_{{\rm r},k}^*{\mathbf{h}}_{{\rm r},k}^T( {\mathbf{I}_N}  - {\mathbf{B}})^H))\widetilde {\mathbf{a}} +  {\mathbf{r}}_{^{5,k}}^T{\mathbf{r}}_{^{5,k}}^* 
   +\frac{1}{2\xi}( 2a-2{{\widetilde{\mathbf{a}}}^H}({{\mathbf{I}}_a}\otimes {\mathbf{A}})\widetilde {\mathbf{a}})$. 

With fixed $\{ {\bf{X}}, {\bf{\Phi }},{\bf{A}},{\bf{B}} \}$, the Lagrangian function of problem \eqref{pro: MUI and waveform discrepancy} with respect to ${\widetilde {\mathbf{a}}}$ is 
\begin{align}\label{pro: Equivalent A}
 \mathop {\min }\limits_{\substack{ \widetilde {\mathbf{a}} }} 
 \; {\mathcal{L}_{\widetilde {\mathbf{a}}}}( {\widetilde {\mathbf{a}},\xi } ) \;\;
 \textrm{s.t.}
 \; \tilde{a}_i \in \{0,1\}, i = 1, 2,\cdots, Na,
 \end{align}
where ${\widetilde{\mathcal{L}}_{\widetilde{\mathbf{a}}}}( {\widetilde {\mathbf{a}}})\! =\! \widetilde {\mathbf{r}}_1^H\widetilde{\mathbf{a}} \!\!+\! \!{\widetilde {\mathbf{a}}^H}{\widetilde {\mathbf{r}}_1} \!+\! {\widetilde{\mathbf{a}}^H}{\mathbf{R}}\widetilde {\mathbf{a}}$, ${\widetilde {\mathbf{r}}_1}\! =\! \rho \sum\nolimits_{k = 1}^K ({{{\mathbf{X}}_{\rm r}}{\mathbf{r}}_{^{5,k}}^*}) \!\!\otimes\!\! ( {( {{\mathbf{I}_N} - {\mathbf{B}}} ){\mathbf{h}}_{{\rm r},k}^*})$, and ${\mathbf{R}} = \rho \sum\nolimits_{k = 1}^K ( {{{\mathbf{X}}_{\rm r}}{\mathbf{X}}_{\rm r}^H}) \otimes ( {( {{\mathbf{I}_N} - {\mathbf{B}}} ){\mathbf{h}}_{{\rm r},k}^*{\mathbf{h}}_{{\rm r},k}^T{{( {{\mathbf{I}} - {\mathbf{B}}})}^H}} ) - \frac{1}{\xi }({{\mathbf{I}}_a} \otimes {\mathbf{A}})$.

The optimization of $\widetilde{\mathbf{A}}$ can be solved following a similar MM-based procedure as in the subproblem w.r.t. $\mathbf{A}$. Therefore, the resulting solution is given by
\begin{equation}\label{equ: opt_tilde a}
{\bar a}_{l,i}\!\! =
\begin{cases}
 1, & \textrm{if}\;i=\!\mathop{\arg\min}\limits_{m}\{{\bf c}[(l-1)N+1:lN]_{m}\}, \\
 0, & \!\textrm{otherwise}, 
\end{cases}
\end{equation}
where ${\bf c} = \Re \left\{{\widetilde{\bf{r}}_2}\right\}$, $l\in\{1,\cdots, a\}$, $i,m\in\{1,\cdots, N\}$, and $\mathop{\arg\min}\limits_{m}\{{\bf c}[(l-1)N+1:lN]_{m}\}$ gives the index corresponding to the minimum value of elements in ${\bf c}[(l-1)N+1:lN]$. In addition, we have $\widetilde{\mathbf{r}}_2
=2\widetilde{\mathbf{r}}_1+2(\mathbf{R} -\mathbf{\Lambda}_2)\widetilde{\mathbf{a}}^{(t)}$ with ${{\mathbf{\Lambda }}_2} = {{\bar \lambda }_{\max }}( {\mathbf{R}} ){{\mathbf{I}}_N}$.
 
It is observed from \eqref{equ: opt_tilde a} that ${\bar a}^{\rm{\mathrm{opt}}}_{l,i} = 1$ holds when the $i$-th element of ${\bf c}[(l-1)N+1:lN]$ corresponds to the minimum value within the $N$ elements of the $l$-th block of ${\bf c}$. Furthermore, the constraint \eqref{con: A tilde} is satisfied according to the resorting algorithm, as shown in Section III-D in \cite{ji2025model}.

Taking into account the penalty term $\xi$, the step size $\eta$ is introduced in the iteration algorithm as follows:
{\small
\begin{align}\label{equ: penalty term}
    \xi^{(t+1)} = \eta \xi^{(t)},
\end{align}}where $\xi^{(t)}$ denotes the penalty term at the $t$-th step.

\subsection{Muting Mode Switching Matrix Optimization}
With fixed $\{ {\bf{X}}, {\bf{\Phi }},{\bf{A}},{\widetilde {\bf{A}}}  \}$, the Lagrangian function of problem \eqref{pro: MUI and waveform discrepancy} with respect to ${\bf{B}}$ is
${\mathcal{L}_{\mathbf{B}}}( {\mathbf{B}})= \rho \sum\nolimits_{k = 1}^K\| {\mathbf{h}}_{d,k}^H{{\mathbf{X}}_b} + {\bm{\phi} ^H}\operatorname{diag}({{\mathbf{h}}_{{\rm r},k}^H}) ({{{\mathbf{I}}_N} - {\mathbf{A}} - {\mathbf{B}}} ){\mathbf{G}}{{\mathbf{X}}_b} + {\mathbf{h}}_{{\rm r},k}^H({{\mathbf{I}_N} - {\mathbf{B}}})\widetilde {\mathbf{A}}{{\mathbf{X}}_{\rm r}} - {\mathbf{s}}_k^T\|_2^2$. 
By defining ${\bf{b}} = \operatorname{diag}({\bf{B}})$, we have 
${\mathcal{L}_{\mathbf{b}}}( {\mathbf{b}} ) = \rho \sum\nolimits_{k = 1}^K {\| {{\mathbf{r}}_{{6,k}}^T + {{\mathbf{b}}^T}{{\mathbf{R}}_{2,k}}} \|_2^2}$,
with ${\mathbf{r}}_{{6,k}}^T = {\mathbf{h}}_{{\rm d},k}^H{{\mathbf{X}}_{\rm b}} + {\phi ^H}\operatorname{diag}( {{\mathbf{h}}_{{\rm r},k}^H} )( {{\mathbf{I}}_N} - {\mathbf{A}}){\mathbf{G}}{{\mathbf{X}}_{\rm b}} + {\mathbf{h}}_{{\rm r},k}^H\widetilde {\mathbf{A}}{{\mathbf{X}}_{\rm r}} - {\mathbf{s}}_k^T$ and ${{\mathbf{R}}_{2,k}} = {\mathbf{\Phi}}\operatorname{diag} ( {{\mathbf{h}}_{{\rm r},k}^H} ){\mathbf{G}}{{\mathbf{X}}_b} + \operatorname{diag}( {{\mathbf{h}}_{{\rm r},k}^H} )\widetilde {\mathbf{A}}{{\mathbf{X}}_{\rm r}}$. After the mathematical manipulations, we have ${\widetilde {\mathcal{L}}_{\mathbf{b}}}( {\mathbf{b}}) = {{\mathbf{b}}^H}{{\mathbf{R}}_3}{\mathbf{b}} + {\mathbf{r}}_7^H{\mathbf{b}} + {{\mathbf{b}}^H}{{\mathbf{r}}_7}$ with ${{\mathbf{r}}_7} =  - \rho \sum\nolimits_{k = 1}^K {{{\mathbf{R}}_{2,k}}{\mathbf{r}}_{{6,k}}^*} $ and ${{\mathbf{R}}_3} = \rho \sum\nolimits_{k = 1}^K {{{\mathbf{R}}_{2,k}}{\mathbf{R}}_{2,k}^*}$. 

Similarly, the optimization of $\widetilde{\mathbf{B}}$ can be solved by an MM-based method. Therefore,
the optimal solution ${\bf{b}}$ is 
{\small
\begin{equation} \label{equ: opt_b}
{b}^{\mathrm{opt}}_{i} =
\begin{cases}
 1, &\Re\left\{{{\bf{r}}_8}\right\}_{[i]} \in \mathcal{V} , \\
 0, &\Re\left\{{{\bf{r}}_8}\right\}_{[i]}  \notin \mathcal{V},
\end{cases}
\end{equation}}where $\mathcal{V}$ is a set consisting of $b$ minimum values of $\Re\left\{{{\bf{r}}_8}\right\}$ with ${{\mathbf{r}}_8} = 2({{\mathbf{R}}_3} - {{\mathbf{\Lambda }}_3}){{\mathbf{b}}^{(t)}} + {{\mathbf{r}}_7}$ and ${{\mathbf{\Lambda }}_3} = {{\bar \lambda }_{\max }}\left( {{{\mathbf{R}}_3}} \right){{\mathbf{I}}_N}$. 
It is observed from \eqref{equ: opt_b} that the corresponding entry of $\bf{b}$ is set to 1 if the corresponding element of $\Re\left\{{{\bf{r}}_8}\right\}$ belongs to $\mathcal{V}$.

The constraint \eqref{con: A*B} ensures that the element cannot be assigned to both the connection and muting modes. If any selected muting element overlaps with the connected elements, it is replaced by the sub-optimal solution corresponding to the index of a larger value of $\Re\left\{{{\bf{r}}_8}\right\}$ until the constraint \eqref{con: A*B} is satisfied. Finally, the above algorithm is summarized in Algorithm \ref{Algo: 1}.
\begin{algorithm}[t]
\footnotesize
    \SetAlgoLined 
	\caption{APDD Algorithm for Joint Waveform and Beamforming Design}
    \label{Algo: 1}
	\KwIn{$K$, $N_{\rm{t}}$, $N$, $a$, $b$, $\mathbf{G}$, ${\bf{h}}_{{\rm{r}},k}$, ${\bf{h}}_{{\rm{d}},k}$, ${{\sigma _k}}$, ${\bf{S}}$, ${\bf{X}}_{0}$, $\rho$, $\xi^{(0)}$, $\eta$ }
	Randomly initialize ${\bf{X}}$, $\bm \varphi$, $\bf{A}$, $\tilde {\bf{A}}$, ${\bf{B}}$\;\label{step: 1 in A1} 
        \Repeat{the convergence is satisfied} 
        {
        Update ${\bf {X}}$ according to \eqref{equ: x_opt_P}\;\label{step: 3 in A1}
        Update $\bm {\varphi}$ according to \eqref{equ: PI p} until the objective function of \eqref{pro: phi PI D_tilde} converges;\label{step: 4 in A1}\\
        Update $\bf \tilde{A}$ according to \eqref{equ: opt_tilde a};\label{step: 8 in A1}\\
        Update $\bf A$ according to \eqref{equ: opt_a};\label{step: 9 in A1}\\
        Update $\mathbf{B}$ according to \eqref{equ: opt_b} ;\\
        Update $\xi$ according to \eqref{equ: penalty term}; \label{step: 11 in A1}
        }
	Return ${\mathbf{X}}^{\mathrm{opt}}$, ${\bf{\Phi}}^{\mathrm{opt}} = \operatorname{diag}(\bm{\varphi}^{\mathrm{opt}})^{H}$,  ${\bf{\tilde A}}^{\mathrm{opt}}$, ${\bf A}^{\mathrm{opt}} = \operatorname{diag}(\bf{a}^{\mathrm{opt}})$, ${\bf B}^{\mathrm{opt}} = \operatorname{diag}(\bf{b}^{\mathrm{opt}})$\;
    \KwOut{${\bf X}$, ${\bf \Phi}$, ${\bf A}$, ${\bf {\tilde A}}$, ${\bf B}$.}
\end{algorithm}
\vspace{-10pt}
\subsection{Convergence and Complexity Analysis}
First, the optimized DF waveform is obtained in closed-form, as shown in step \ref{step: 3 in A1}. For the passive beamforming, the subproblem for updating ${\bm{\varphi}}$ is optimally solved, which guarantees the maximization of the objective function in \eqref{pro: phi PI D_tilde} and is equivalent to \eqref{pro: MUI and waveform discrepancy}, as shown in step \ref{step: 4 in A1}. Moreover, the objective value of problem \eqref{pro: a} is non-increasing under the MM method, while the convergence of problems \eqref{pro: Equivalent A} and muting mode switching optimization is also ensured by the MM framework in steps \ref{step: 5 in A2}-\ref{step: 7 in A2}. Therefore, Algorithm \ref{Algo: 1} is guaranteed to converge.

In addition, the computational complexity of Algorithm \ref{Algo: 1} is analyzed as follows. Specifically, for the waveform optimization, the computational complexity is dominated by the matrix inversion in \eqref{equ: x_opt_P}, where the matrix dimension is $L(N_{\rm t}+a) \times L(N_{\rm t}+a)$, leading to a complexity of $\mathcal{O}((L(N_{\rm t}+a))^3)$.
For the phase shift optimization, the construction of the matrix $\mathbf{D}$ dominates the complexity, which is given by $\mathcal{O}(KLN^2)$. In addition, problem \eqref{pro: phi PI D_tilde} is solved by the power iteration, and the corresponding complexity is $\mathcal{O}(I_{\rm in} N^2)$, where $I_{\rm in}$ denotes the number of inner iterations. 
For the connection mode switching matrix optimization, the computational complexity mainly arises from the construction of $\widetilde{\mathbf{E}}$ and the computation of its largest eigenvalue, which is given by $\mathcal{O}(KLN^2 + N^3)$. For the equivalent connection mode switching, the complexity is dominated by the construction of $\mathbf{R}$ and its eigenvalue decomposition, which is given by $\mathcal{O}((Na)^3)$. For the muting mode switching, the complexity mainly comes from the construction of $\mathbf{R}_3$ and the computation of its largest eigenvalue, yielding a complexity of $\mathcal{O}(KLN^2 + N^3)$. 
Therefore, the total computational complexity of Algorithm \ref{Algo: 1} is given by $\mathcal{O}(I_{\rm out}((L(N_{\rm t}+a))^3 + KLN^2 + I_{\rm in} N^2 + (Na)^3 + N^3))$, where $I_{\rm out}$ denotes the number of outer iterations.

\section{Deep Unfolding APDD-Net}\label{sec: APDD-Net}
In this section, we propose an APDD-Net by deeply unfolding the APDD algorithm to improve the system performance and the convergence speed.
\begin{figure*}[t] 
 \centering
 \includegraphics[width=0.8\textwidth]{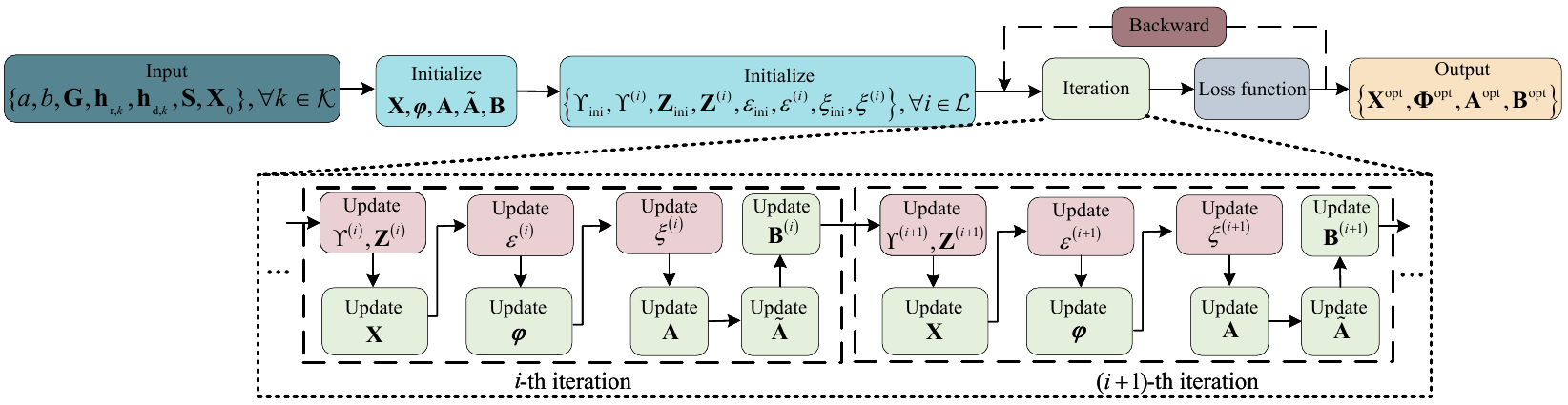}
  \vspace{-10pt}
 \caption{An illustration of the network structure for APDD-Net.}
 \label{fig: APDD-Net}
\vspace{-10pt}
\end{figure*}
\subsection{Proposed Network}
Note that the optimal solution of the waveform design is obtained by the complex matrix inversion operation based on \eqref{equ: x_opt_P}, which suffers from the issue of high computational complexity. To this end, an approximation approach by unfolding a discrete dynamic system is applied for matrix inversion \cite{matrix_inversion_2009, matrix_inversion_2025}.
Specifically, let $\widetilde{\bf{C}}={{\rho \sum\nolimits_{k = 1}^K {{\mathbf{C}}_k^H{{\mathbf{C}}_k}}  + ( {1 - \rho} ){{\mathbf{I}}_{L({N_{\rm t}} + a)}}}} \in \mathbb{C}^{L(N_{\rm t} + a)\times L(N_{\rm t} + a)}$, the inverse of matrix $\widetilde{\bf{C}}$ is given by
\begin{equation}
    {\bm{\Gamma}}(\widetilde{\bf{C}}; {\bm{\Upsilon}}, \mathbf{Z} ) = \bar{\bf{C}} - \bar{\bf{C}}{\bm{\Upsilon}}\operatorname{tanh}(\widetilde{\bf{C}}\bar{\bf{C}}-{\mathbf{I}}_{L({N_{\rm t}} + a)}) + {\bf{Z}},
\end{equation}
with $\bar{\bf{C}} = (\operatorname{diag}(\widetilde{\bf{C}}))^{-1}$, where ${\bm{\Upsilon}}$ and ${\bf{Z}}$ are the trainable parameters, and $\operatorname{diag}(\widetilde{\bf{C}})$ denotes the diagonal matrix of $\widetilde{\bf{C}}$. Then, the network will be designed and trained to predict the suitable trainable parameters, so as to reconstruct the inverse of the matrix. Therefore, the waveform in the $(\ell+1)$-th outer iteration is 
\begin{equation}\label{equ: update X in APDD-Net}
  \small  
\begin{aligned}
    \mathbf{x}^{(\ell+1)}\!\! = \!\!\frac{\sqrt{P_{tot}}{\bm{\Gamma}}(\widetilde{\bf{C}}^{(\ell)};\!{\bm{\Upsilon}}^{(\ell)},\! \mathbf{Z}^{(\ell)}) (\rho\!\!\sum\nolimits_{k=1}^{K}\!{{(\bf{C}}^{(\ell)}_k)^{H}}{\bf{s}}^{(\ell)}_k\!\!+\!\!(1\!\!-\!\!\rho){\bf{x}}^{(\ell)}_0) }{||{\bm{\Gamma}}(\widetilde{\bf{C}}^{(\ell)};\!{\bm{\Upsilon}}^{(\ell)}, \mathbf{Z}^{(\ell)}) (\rho\!\!\sum\nolimits_{k=1}^{K}{{(\bf{C}}^{(\ell)}_k)^{H}}{\bf{s}}^{(\ell)}_k\!\!+\!\!(1\!\!-\!\!\rho){\bf{x}}^{(\ell)}_0)||_{2},}
\end{aligned}
\end{equation}
where ${\bm{\Gamma}}(\widetilde{\bf{C}}^{(\ell)};{\bm{\Upsilon}}^{(\ell)}, \mathbf{Z}^{(\ell)}) = \bar{\bf{C}}^{(\ell)} - \bar{\bf{C}}^{(\ell)}{\bm{{\bm{\Upsilon}}}}^{(\ell)}\operatorname{tanh}(\widetilde{\bf{C}}^{(\ell)}\bar{\bf{C}}^{(\ell)}-{\mathbf{I}}_{L({N_{\rm t}} + a)}) + {\bf{Z}}^{(\ell)}$.

Note that the regularization term $\varepsilon$ determines the convergence speed of inner iterations for passive beamforming design. When the value of $\varepsilon$ is too large, the inner iteration has a very slow convergence speed. Moreover, the inner iteration fails to converge when the positive definiteness of $\widetilde{\bf{D}} + \varepsilon {\bf{I}}_{N+1}$ is not guaranteed. However, with the model-driven NN, the number of inner iterations is set to 1, which can significantly reduce the computational complexity. 

On the other hand, the penalty term $\xi$ is introduced to satisfy the constraint \eqref{con: A +A tilde}. Specifically, at the beginning of the iterations, the term $\xi$ should be judiciously selected such that the first terms in ${\mathcal{L}_{\mathbf{A}}}( {{\mathbf{A}},\xi })$ and ${\mathcal{L}_{\widetilde {\mathbf{A}}}}( {\widetilde {\mathbf{A}},\xi })$ dominate the value.
The goal of the corresponding objective function is to minimize the first term. When the value of the first term tends to be stable, the $\xi$ is selected to ensure that the latter term dominates the value of the corresponding objective function. Therefore, the adaptive penalty term $\xi$ is desirable during iterations. 

Based on the above discussions, the APDD-Net is shown in Fig. \ref{fig: APDD-Net}.
First, we set $\{ {\bm{\Upsilon}}_{\rm{ini}}, {\bm{\Upsilon}}^{(\ell)}, {\bf{Z}}_{\rm{ini}},{\bf{Z}}^{(\ell)}, \varepsilon_{\rm{ini}}, \varepsilon^{(\ell)}, \xi_{\rm{ini}},\xi^{(\ell)}\}$ as the trainable parameters. In addition, the iterations of Algorithm \ref{Algo: 1} are deeply unfolded into a layer-wise NN. The input of APDD-Net is the channels and weighted factor, and the output is the optimized variables, i.e., $\{{\bf X}, {\bf \Phi}, {\bf A}, {\bf {\tilde A}}, {\bf B}\}$. The maximum number of iterations is fixed and determined by the APDD algorithm, where $I'_{\rm out}$ denotes the number of iterations for the APDD-Net algorithm with $\mathcal{L} = \{1, 2, \cdots, I'_{\rm out}\}$.
The whole APDD-Net algorithm is summarized in Algorithm \ref{Algo: 2}.

\subsection{Complexity Analysis}
In addition, for the proposed APDD-Net algorithm, the computational complexity is analyzed as follows. Since the APDD-Net is constructed by unfolding the APDD iterations, each layer corresponds to one iteration of the APDD algorithm.
Specifically, for the waveform optimization, the closed-form solution in \eqref{equ: x_opt_P} is replaced by a linear transformation with trainable parameters. Therefore, the corresponding complexity results from matrix multiplications, i.e., $\mathcal{O}((L(N_{\rm t}+a))^3)$.
Although the computational complexity of the exact inverse and the matrix inverse approximation is theoretically comparable, the exact inverse entails more intricate operations, such as matrix triangularization and back substitution, resulting in higher practical computational overhead \cite{matrix_inversion_2025}. Consequently, APDD-Net achieves lower complexity than APDD for waveform optimization at each iteration.
For the phase shift optimization, the construction of the matrix $\mathbf{D}$ still dominates the computational complexity, which is given by $\mathcal{O}(KLN^2)$. Due to the trainable regularization term, the number of power iterations is reduced to one, resulting in a complexity of $\mathcal{O}(N^2)$.
In addition, the number of outer iterations is reduced to $I'_{\rm out}$ with the trainable penalty term, where $I'_{\rm out} \ll I_{\rm out}$.
Therefore, the total computational complexity of the APDD-Net algorithm is given by
$\mathcal{O}(I'_{\rm out}((L(N_{\rm t}+a))^3 + KLN^2 +  N^2 + (Na)^3 + N^3))$.
\begin{algorithm}[t]
\footnotesize
    \label{Algo: 2}
    \SetAlgoLined 
	\caption{APDD-Net Algorithm for Joint Waveform and Beamforming Design}
	\KwIn{$K$, $N_{\rm{t}}$, $N$, $a$, $b$, $\mathbf{G}$, ${\bf{h}}_{{\rm{r}},k}$, ${\bf{h}}_{{\rm{d}},k}$, ${{\sigma _k}}$, ${\bf{S}}$, ${\bf{X}}_{0}$, $\rho$}
	Randomly initialize ${\bf{X}}$, $\bm \varphi$, $\bf{A}$, $\tilde {\bf{A}}$, ${\bf{B}}$\;\label{step: 1 in A2} 
        Initialize trainable variables \{${\bm{\Upsilon}}_{\rm{ini}}$, ${\bm{\Upsilon}}^{(i)}$, $\mathbf{Z}_{\rm{ini}}$, $\mathbf{Z}^{(i)}$, $\xi_{\rm{ini}}$, $\xi^{(i)}$, $\varepsilon_{\rm{ini}}$, $\varepsilon^{(i)}$\}, $\forall i \in \mathcal{L}$\;
        \Repeat{the convergence is satisfied} 
        {
        Update ${\bf{X}}$ according to \eqref{equ: update X in APDD-Net} and trainable variables ${\bm{\Upsilon}}^{(i)}, \mathbf{Z}^{(i)}$\;\label{step: 4 in A2}
        Update $\bm {\varphi}$ according to \eqref{equ: PI p} and trainable variable $\varepsilon^{(i)}$;\label{step: 5 in A2}\\
        Update $\bf \tilde{A}$ according to \eqref{equ: opt_tilde a} and trainable variable $\xi^{(i)}$;\label{step: 6 in A2}\\
        Update $\bf A$ according to \eqref{equ: opt_a} and trainable variable $\xi^{(i)}$;\label{step: 7 in A2}\\
        Update $\mathbf{B}$ according to \eqref{equ: opt_b};\label{step: 8 in A2}
        }
	Return ${\mathbf{X}}^{\mathrm{opt}}$, ${\bf{\Phi}}^{\mathrm{opt}} = \operatorname{diag}(\bm{\varphi}^{\mathrm{opt}})^{H}$,  ${\bf{\tilde A}}^{\mathrm{opt}}$, ${\bf A}^{\mathrm{opt}} = \operatorname{diag}(\bf{a}^{\mathrm{opt}})$, ${\bf B}^{\mathrm{opt}} = \operatorname{diag}(\bf{b}^{\mathrm{opt}})$\;
    \KwOut{${\bf X}$, ${\bf \Phi}$, ${\bf A}$, ${\bf {\tilde A}}$, ${\bf B}$.}
\end{algorithm}

\section{Simulation Results}\label{sec: simulation}
In this section, the simulation results are presented to verify the effectiveness of the proposed algorithms. If not specified, we set $N_{\rm{t}} =16$, $K = 6$, $L = 28$, $N = 50$, $P_{\rm{tot}} = 30$ dBm, $\rho = 0.5$, $f_{0} =25$ GHz, $f_{\rm{s}} =60$ GHz, 
$b=2$, $a = 10$, $\sigma^2 = \sigma_{{\rm s}}^2=-90$ dBm, and $\upsilon=0.1$. Under the 3D Cartesian coordinate system, the ISAC-BS and RDARS are located at (0, 0, 15) m and (0, 10, 15) m, respectively. The UEs are randomly distributed within a circle, where its center and radius are (10, 50, 1.5) m and 5 m, respectively. The target is located at (50, 10, 20) m. The trainable matrices, i.e., ${\bm{\Upsilon}}_{\rm{ini}}$ and ${\bf{Z}}_{\rm{ini}}$, are randomly initialized.
The main simulation parameters are summarized in Table \ref{tb:listbridges}.

\begin{table}[t]
\scriptsize
\centering 
\begin{threeparttable}
\caption{The Main Simulation Parameters.}
\label{tb:listbridges}
\begin{tabular}{llll}
\toprule
\textbf{Description} & \textbf{Parameter} & \textbf{Value} \\
\midrule
Path loss exponent of channel ${\bf{G}}$&  & 2.2 \\
Path loss exponent of channel ${\bf{h}}_{{\rm{r}},k}$& & 2.4 \\
Path loss exponent of channel ${\bf{h}}_{{\rm{d}},k}$&  & 2.2 \\
Path loss exponent of channel ${\bf{h}}_{{\rm{dt}}}$& $\alpha_{\mathrm{bt}}$ & 2.1 \\
Path loss exponent of channel ${\bf{h}}_{{\rm{rt}}}$& $\alpha_{\mathrm{rt}}$ & 2.2 \\
Rician factor $\tau$ & 10 \\
Penalty term in the initialization step& $\xi^{\rm ini}$ & $10^{6}$ \\
Regularization term in the initialization step& $\varepsilon^{\rm ini}$ & $10^{-8}$ \\
Step size& $\eta$ & 0.1 \\
Number of iterations&  & 15 \\
Number of channel realizations&  & 5000\\
\bottomrule
\end{tabular}
\label{tab: 2}
\end{threeparttable}
\vspace{-10pt}
\end{table}


By considering the general Rician fading channel model, we have
${\bf{G}} = \kappa_{{\rm{b}}}(\sqrt{\frac{\tau}{\tau+1}} \overline{\bf{G}} + \sqrt{\frac{1}{\tau+1}} \widetilde{{\bf{G}}})$,
 ${\bf{h}}_{{\rm{r}},k} = \kappa_{{\rm{r}},k} (\sqrt{\frac{\tau}{\tau+1}}\overline{{\bf{h}}}_{{\rm{r}},k} + \sqrt{\frac{1}{\tau+1}}\widetilde{{\bf{h}}}_{{\rm{r}},k})$, and ${\bf{h}}_{{\rm{d}},k} = \kappa_{{\rm{d}},k} (\sqrt{\frac{\tau}{\tau+1}}\overline{{\bf{h}}}_{{\rm{d}},k} + \sqrt{\frac{1}{\tau+1}}\widetilde{{\bf{h}}}_{{\rm{d}},k})$,
where $\tau$ denotes the Rician factor.
The elements of the non-line-of-sight (NLoS) components, denoted by $\widetilde{{\bf{G}}}$, $\widetilde{\bf{h}}_{{\rm r},k}$, and $\widetilde{{\bf{h}}}_{{\rm{d}},k}$, are characterized by a standard complex Gaussian distribution. The LoS components of BS-RDARS, RDARS-UE $k$, and BS-UE $k$ channels are represented by $\overline{\bf{G}} $, $\overline{{\bf{h}}}_{{\rm{r}},k}$, and $\overline{{\bf{h}}}_{{\rm{d}},k}$, respectively.

For comparison, the following schemes are considered: \textbf{1) DAS}: The number of BS antennas and distributed antennas are $\tilde{N}_{\rm{t}} = N_{\rm{t}}$ and $\tilde{N}_{\rm{dt}} = a$, respectively. The corresponding waveform is obtained by the APDD algorithm; \textbf{2) RIS}: For the RIS-assisted system, the number of BS antennas and passive elements are $\bar{N}_{\rm t} = N_{\rm{t}} + a$ and $\bar{N} = N$, respectively. The APDD algorithm is applied to optimize the waveform and passive beamforming; \textbf{3) RDARS APDD}: For the RDARS-aided system, the waveform design, passive beamforming, and dynamic element configuration are obtained by the APDD algorithm; \textbf{4) RDARS APDD-Net}: The APDD-Net algorithm is applied to the RDARS-aided system by unfolding the iterations of the APDD algorithm into a neural network with trainable parameters.

\begin{figure}[t] 
 \centering
 \includegraphics[width=0.38\textwidth]{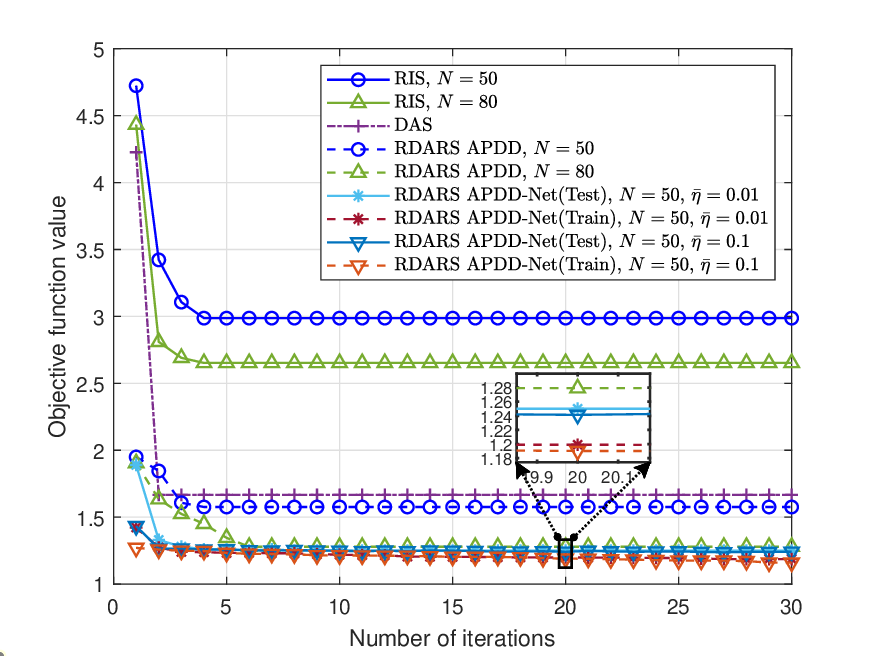}
  \vspace{-10pt}
 \caption{Convergence behavior of different schemes in terms of the objective value versus the number of iterations. $\tilde{\eta}$ denotes the learning rate used in the APDD-Net.}
 \label{fig: convergence}
 \vspace{-10pt}
\end{figure}

Fig. \ref{fig: convergence} shows the objective function value versus the number of iterations for different architectures. 
It is first observed that the objective values of proposed algorithms among different architectures tend to converge as the number of iterations increases, which demonstrates the effectiveness of proposed algorithms.
It is also observed that the RDARS-aided architecture outperforms the DAS and RIS-aided systems, which is mainly attributed to the selection gain brought by the flexible element configuration. Specifically, the flexible adjustment of the RDARS transmit elements' positions facilitates the adaptation to the varying channel environment. In addition, the MUI can be mitigated by optimizing the placement of RDARS muting elements, such that the objective function mainly focuses on waveform design.

Furthermore, the APDD-Net algorithm has a faster convergence speed compared to the APDD algorithm. This is expected due to the deep unfolding iterations with trainable parameters, which make the penalty and regularization terms adaptive across iterations.
Besides, the approximate matrix inversion relying on the trainable parameters achieves a comparable performance with low complexity, which demonstrates the effectiveness of the model-driven NN. 
By comparing the performance of the trained and testing APDD-Nets, it can be found that the performance gap is very small, revealing the robustness of the model-driven APDD-Net algorithm. 
This is because the NN deeply unfolds the iterations of the APDD algorithm by integrating several trainable parameters. 
The performance gap caused by varying channel conditions is mitigated through the iterative updates with trainable parameters, which enhances the adaptability of the NN.
In addition, the performance of the APDD-Net with $N = 50$ even outperforms the APDD with $N = 80$, which stems from the capability of the model-driven method to escape from local optima.

\begin{figure}[t] 
 \centering
 \includegraphics[width=0.38\textwidth]{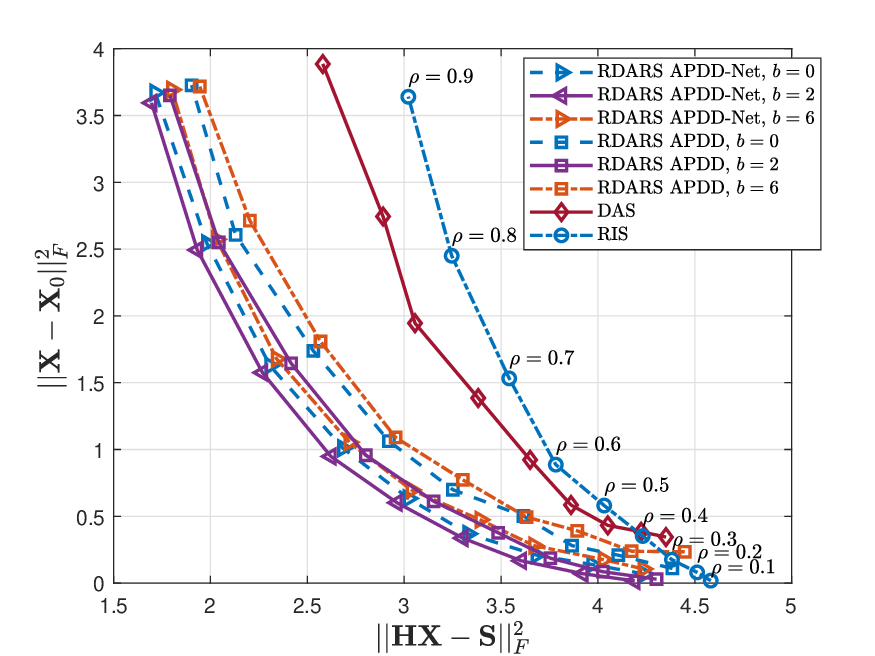}
 \vspace{-10pt}
 \caption{{Performance trade-off between the MUI energy and waveform discrepancy.}}
 \label{fig: rho_Wave_MUI}
 \vspace{-10pt}
\end{figure}

Fig. \ref{fig: rho_Wave_MUI} shows the performance trade-off between the MUI energy and waveform discrepancy for the proposed algorithms.
It is observed that the waveform discrepancy decreases with the MUI energy for all architectures and algorithms due to the limited spatial DoF, i.e., a trade-off exists between sensing and communication performance.
It is also observed that the APDD-Net algorithm achieves a better performance compared to the APDD. This is because the APDD-Net can learn the mapping between the channel state information and the optimization variables.
Specifically, the exact matrix inversion related to the DF waveform is approximated by matrix multiplication in the model-driven NN. Moreover, the regularization term predicted by the APDD-Net guarantees the positive definiteness of $\widetilde {\mathbf{D}} + \varepsilon {{\mathbf{I}}_{N + 1}}$. 
In addition, the APDD-Net generates the adaptive penalty term to satisfy the integer constraints. 
Moreover, a moderate number of muting elements improves the system performance, thereby verifying the effectiveness of the muting mode.
The above results demonstrate the superiority of the model-driven APDD-Net and highlight the potential of integrating the RDARS into ISAC systems.

\begin{figure}[t]
 \centering
 \includegraphics[width=0.38\textwidth]{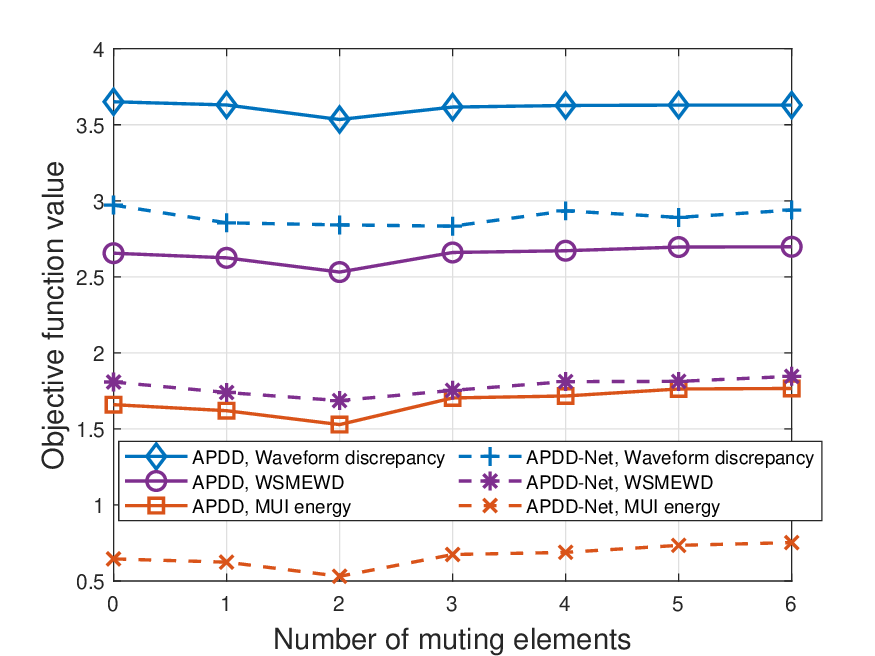}
 \vspace{-10pt}
 \caption{Objective function value versus the number of muting elements.}
 \label{fig: weight_of}
 \vspace{-10pt}
\end{figure}

In Fig. \ref{fig: weight_of}, we plot the objective value versus the number of muting elements. It is observed that APDD-Net achieves a better communication and sensing performance than APDD. 
Specifically, the waveform discrepancies between APDD and APDD-Net algorithms initially decrease with the number of muting elements, which is expected since the MUI is mitigated, thus facilitating the waveform design to improve sensing performance. As the number of muting elements increases, the passive beamforming gain decreases, and the main focus of the waveform design is to reduce the MUI. 
Therefore, the number of muting elements presents a trade-off between the MUI energy and waveform discrepancy for the APDD algorithm.
By contrast, the MUI energy and waveform discrepancy exhibit the same trend for the APDD-Net algorithm, which demonstrates the superiority of the model-driven method. 
This is because the trained network can well extract the characteristics of multiple channel realizations.
In addition, the waveform discrepancy of APDD-Net is smaller than that of the APDD algorithm when $b$ is small, which further demonstrates the effectiveness of approximate matrix inversion. 

\begin{figure}[t]
 \centering
 \includegraphics[width=0.38\textwidth]{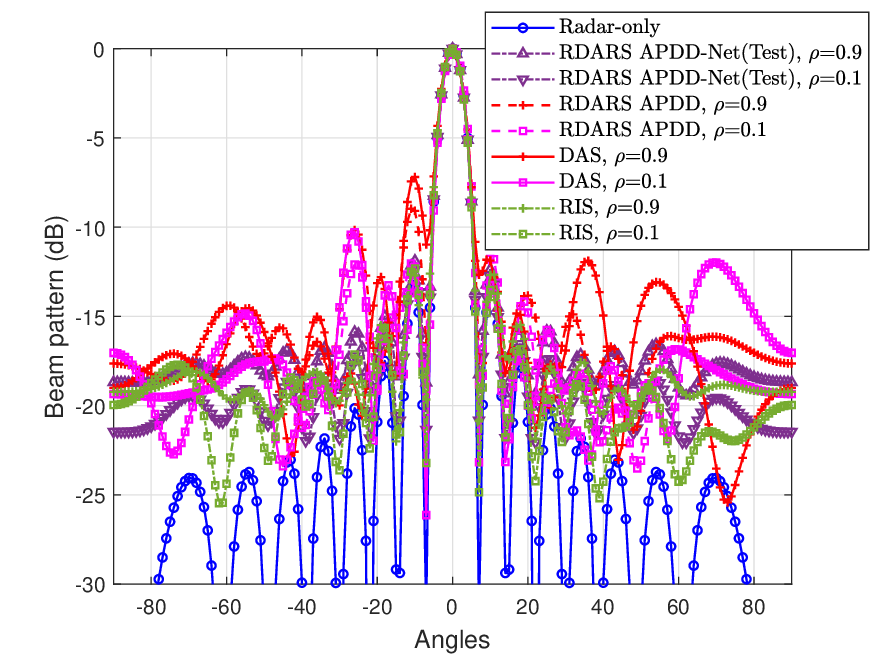}
 \vspace{-10pt}
 \caption{Beam pattern versus angles with different $\rho$ where $N_{t} = 16$.}
 \label{fig: beampattern}
 \vspace{-10pt}
\end{figure}

In Fig. \ref{fig: beampattern}, by setting the reference beam pattern for the ISAC BS as 
$G(\theta) = \frac{1}{N_{\rm{t}}}|\sqrt{\frac{(N_{\rm{t}} + a)L}{ P_{\rm{tot}}} } \mathbf{c}^{H}(\theta) \mathbf{c}(\theta_{\rm{b}}) {\tilde{\bf{x}}}|$ \cite{zhang2025joint_deep_unfolding}, we plot the beam pattern of the waveform transmitted from the ISAC BS. 
The beam pattern of the radar-only scheme is designed by considering only the sensing task, without taking communication users into account.
One can observe that the designed waveform becomes closer to the reference waveform as the weighting factor decreases. This is because the objective function mainly aims to minimize the waveform discrepancy with a small $\rho$. 
In addition, the waveform for the DAS exhibits a high sidelobe, which may result in a relatively poor sensing performance. 
The reason is that the distributed antenna structure in DAS leads to reduced spatial coherence in beamforming, resulting in higher sidelobe levels.

\begin{figure}[t] 
 \centering
 \includegraphics[width=0.38\textwidth]{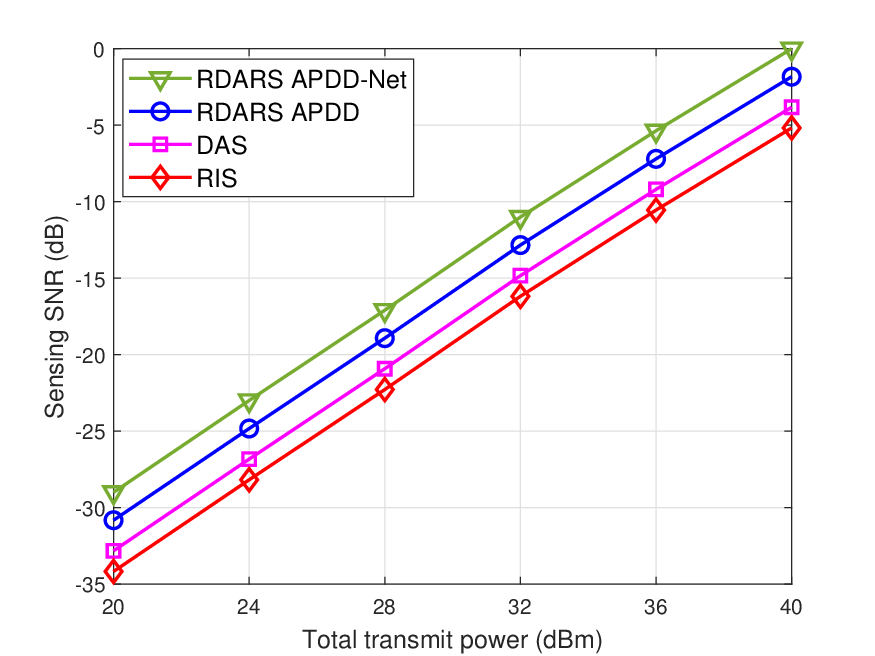}
  \vspace{-10pt}
 \caption{Sensing SNR versus the total transmit power.}
 \label{fig: sensing SNR}
  \vspace{-10pt}
\end{figure}

Fig. \ref{fig: sensing SNR} compares the sensing SNR for different architectures with the APDD and APDD-Net algorithms.
It is observed that the sensing SNR increases with the total transmit power due to the increased echo power. Moreover, we observe that the RDARS architecture outperforms the DAS. This is because the passive elements create the additional reflection link, resulting in the additional passive beamforming gain compared to the DAS. 
In addition, the dynamically connected elements bring the distributed gain compared with the RIS-aided system, and thus a superior performance can be achieved for the RDARS architecture. 
Furthermore, the APDD-Net achieves a higher sensing SNR than the APDD algorithm, as the model-driven NN provides a better trade-off between sensing and communication performance. 
The above result further demonstrates the potential of model-driven NNs in ISAC systems.

\begin{figure}[t]
 \centering
 \includegraphics[width=0.38\textwidth]{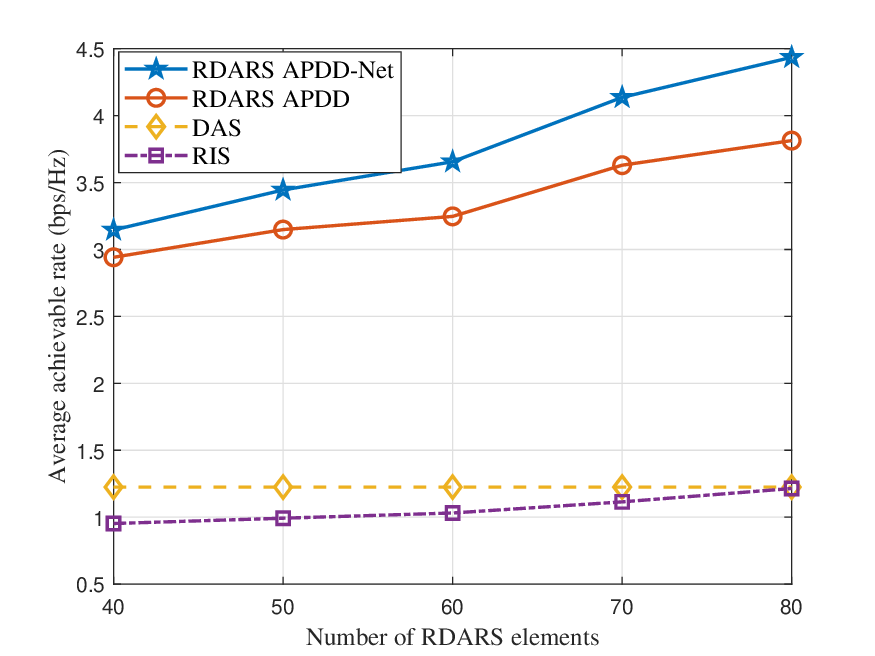}
 \vspace{-10pt}
  \caption{Average achievable rate versus the total number of RDARS elements.}
 \label{fig: rate VS N}
 \vspace{-10pt}
\end{figure}

Fig. \ref{fig: rate VS N} compares the average achievable rate versus the number of RDARS elements for different architectures.
It is observed that the average achievable rate increases with the number of passive elements for the RDARS and RIS-aided systems, since the passive beamforming gain brought by the passive elements increases. It is also observed that the DAS attains only a 0.86\% rate improvement compared to the RIS-aided system when $N = 80$. This indicates that the passive elements can provide a comparable beamforming gain in a cost-effective manner. 
In addition, we observe that the APDD-Net is superior to the APDD, since the model-driven deep learning overcomes local optima and improves communication performance.


\begin{figure}[t] 
 \centering
 \includegraphics[width=0.38\textwidth]{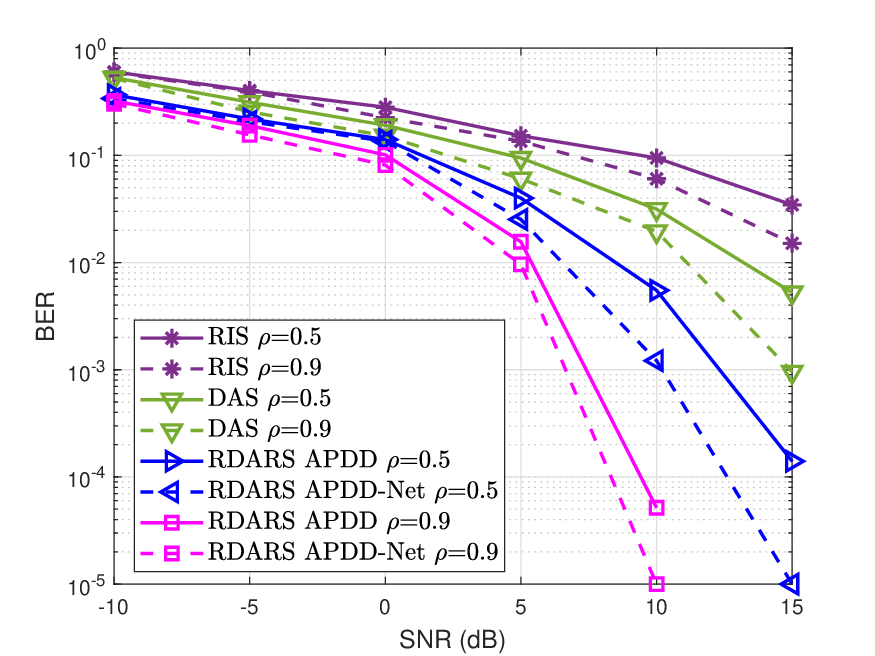}
  \vspace{-10pt}
 \caption{BER versus SNR.}
 \label{fig: ber}
 \vspace{-15pt}
\end{figure}
Fig. \ref{fig: ber} shows the BER versus the transmit SNR for the proposed algorithms. It is observed that the BER decreases with SNR for all architectures and algorithms, as expected. It is also observed that the BER decreases with the weighting factor since the variables are optimized to minimize the MUI in this case. Moreover, the RDARS architecture attains a lower BER compared to the DAS and RIS-aided systems. In addition, the model-driven APDD-Net achieves a lower BER compared to the APDD algorithm, especially at high SNR. 
This is because the model-driven NN captures a better mapping function from the channels to the passive beamforming and mode switching matrices.

{\section{Conclusion}\label{sec: conclusion}}
In this paper, we considered a RDARS-aided ISAC system by introducing a new muting mode, where the weighted sum of the MUI energy and waveform discrepancy was minimized by jointly optimizing the DF waveform, passive beamforming, and the connection and muting mode switching matrices for RDARS. 
We first verified the muting gain introduced by muting elements.
Subsequently, the APDD algorithm was proposed to solve the MINLP problem by utilizing the alternating optimization, penalty dual decomposition, power iteration, and MM algorithms.
Moreover, the parameters determining the computational complexity of the APDD algorithm and system performance were analyzed, and we proposed a model-driven deep learning network by deeply unfolding the APDD algorithm into a layer-wise network, i.e., APDD-Net, with these parameters being trained.
Simulation results demonstrated the effectiveness of mode switching, and highlighted the superiority of deep unfolding in jointly optimizing the waveform and mode switching for the considered system.

\appendices
{\section*{Appendix A: Proof of Proposition \ref{Propo: 1}} \label{appendix: A}}
With the MRT beamforming scheme, 
the interference power caused by UE $j$ at UE $k$ is given by $I_{k,j}  = P_j|\mathbf h_k^H\mathbf w_j|^2 = P_j \frac{|\mathbf h_k^H\mathbf h_j|^2}{\|\mathbf h_j\|_2^2}$.
Then, the equivalent channel of UE $k$ is decomposed as $\mathbf h_k=\mathbf h_{0,k}+\mathbf h_{{\rm ref},k}$, where $\mathbf h_{0,k}$ denotes the non-reflection channel component related to the BS-UE and RDARS-UE links, and $\mathbf h_{{\rm ref},k}$ denotes the reflection channel component generated by the passive reflecting elements. Similarly, we have 
$\mathbf h_j=\mathbf h_{0,j}+\mathbf h_{{\rm ref},j}$.

According to the definitions of $E_k$ and $G_k$, it follows that 
$\|\mathbf h_{0,k}\|_2^2 = E_k$, and $\|\mathbf h_{{\rm ref},k}\|_2^2= N^2_{\rm p}G_k$.
Then, the channel correlation of UEs $k$ and $j$ is
$\mathbf h_k^H\mathbf h_j
=\mathbf h_{0,k}^H\mathbf h_{0,j}+\mathbf h_{0,k}^H\mathbf h_{{\rm ref},j}+\mathbf h_{{\rm ref},k}^H\mathbf h_{0,j}
+\mathbf h_{{\rm ref},k}^H\mathbf h_{{\rm ref},j}$.
Let $A_{k,j}
\triangleq
\mathbf h_{{\rm ref},k}^H\mathbf h_{{\rm ref},j}$ represent the correlation caused by the reflection link correlation coefficient and $B_{k,j}
\triangleq
\mathbf h_{0,k}^H\mathbf h_{0,j}
+
\mathbf h_{0,k}^H\mathbf h_{{\rm ref},j}
+
\mathbf h_{{\rm ref},k}^H\mathbf h_{0,j}$ represent the correlation caused by the non-reflection components and the mixed components.
By invoking the triangle inequality, we have $|B_{k,j}|
\le|\mathbf h_{0,k}^H\mathbf h_{0,j}|+|\mathbf h_{0,k}^H\mathbf h_{{\rm ref},j}|+|\mathbf h_{{\rm ref},k}^H\mathbf h_{0,j}|$.
Moreover, with the Cauchy-Schwarz inequality, we have
$|\mathbf h_{0,k}^H\mathbf h_{0,j}|
\le\|\mathbf h_{0,k}\|_2\|\mathbf h_{0,j}\|_2=\sqrt{E_kE_j}$, $|\mathbf h_{0,k}^H\mathbf h_{{\rm ref},j}| \le\|\mathbf h_{0,k}\|_2\|\mathbf h_{{\rm ref},j}\|_2=N_{\rm p}\sqrt{E_kG_j}$,
and $|\mathbf h_{{\rm ref},k}^H \mathbf h_{0,j}|\le\|\mathbf h_{{\rm ref},k} \|_2\|\mathbf h_{0,j}\|_2=N_{\rm p}\sqrt{G_kE_j}$.
Therefore, it follows that 
$|B_{k,j}|\le\sqrt{E_kE_j}+N_{\rm p}\sqrt{E_kG_j}+N_{\rm p}\sqrt{G_kE_j}$.
On the other hand, the reflection link correlation value is defined as
$|A_{k,j}|^2=|\mathbf h_{{\rm ref},k}^H \mathbf h_{{\rm ref},j} |^2=N^4_{\rm p}C_{k,j}$.
Then, we have 
$|A_{k,j}|=N^2_{\rm p}\sqrt{C_{k,j}}$.
Therefore, when the condition $N^2_{\rm p}\sqrt{C_{k,j}} \gg \sqrt{E_kE_j} + N_{\rm p}\sqrt{E_kG_j} + N_{\rm p}\sqrt{G_kE_j}$ holds, we have $|A_{k,j}|\gg |B_{k,j}|$, which implies reflection-link correlation dominates the non-reflection and mixed components.

Under the above sufficient condition, the MUI under MRT is mainly caused by the common reflected components. The interference term can be approximated as $P_j\frac{|\mathbf h_k^H \mathbf h_j |^2}{\|\mathbf h_j \|_2^2}\approx P_j\frac{N^4_{\rm p}C_{k,j}}{E_j+N^2_{\rm p}G_j}$.
This completes the proof.

{\section*{Appendix B: Proof of Proposition \ref{Propo: 2}} \label{appendix: B}}
After some mathematical manipulations, we have
$\gamma_k({\tilde{N}_{\rm p}})=
\frac{
P_k(E_k+{\tilde{N}_{\rm p}}G_k)(E_j+{\tilde{N}_{\rm p}}G_j)
}{
\sigma^2(E_j+{\tilde{N}_{\rm p}}G_j)+P_jC_{k,j}{\tilde{N}^2_{\rm p}}
}$.
Let $\mathcal N_k({\tilde{N}_{\rm p}})=(E_k+{\tilde{N}_{\rm p}}G_k)(E_j+{\tilde{N}_{\rm p}}G_j)$
and
$\mathcal D_k({\tilde{N}_{\rm p}})=\sigma^2(E_j+{\tilde{N}_{\rm p}}G_j)+P_jC_{k,j}{\tilde{N}^2_{\rm p}}$, we have
$\gamma_k ({\tilde{N}_{\rm p}})=P_k\frac{\mathcal N_k({\tilde{N}_{\rm p}})}{\mathcal D_k({\tilde{N}_{\rm p}})}$. 
Accordingly, 
we have
$\frac{d\gamma_k({\tilde{N}_{\rm p}})}{d{\tilde{N}_{\rm p}}}=\frac{P_k\Omega_k({\tilde{N}_{\rm p}})}{(\sigma^2(E_j+{\tilde{N}_{\rm p}}G_j)+P_jC_{k,j}{\tilde{N}^2_{\rm p}})^2}$, where $\Omega_k({\tilde{N}_{\rm p}})=G_k\sigma^2(E_j+{\tilde{N}_{\rm p}}G_j)^2 -
P_jC_{k,j}{\tilde{N}_{\rm p}}(2E_kE_j+(E_kG_j+G_kE_j){\tilde{N}_{\rm p}})$.
Next, $\Omega_k({\tilde{N}_{\rm p}})$ can be expanded as
$\Omega_k({\tilde{N}_{\rm p}})=c_{k,2}{\tilde{N}^2_{\rm p}}+c_{k,1}{\tilde{N}_{\rm p}}+c_{k,0}$,
where $c_{k,0}=G_k\sigma^2E_j^2>0$ and $c_{k,1}=2G_k\sigma^2E_jG_j-2P_jC_{k,j}E_kE_j$, and $c_{k,2}
=G_k\sigma^2G_j^2-P_jC_{k,j}(E_kG_j+G_kE_j)$.
Therefore, the optimal $\gamma_k({\tilde{N}_{\rm p}})$ is given by
${\tilde{N}^{\rm opt}_{{\rm p}, k}}=\frac{-c_{k,1}-\sqrt{c_{k,1}^2-4c_{k,2}c_{k,0}}}{2c_{k,2}}$. Recalling that $\tilde N_{\rm p}=(N-a-b)^2$, we have $b_k^{\rm opt}=\operatorname{round}(N-a-\sqrt{{\tilde{N}^{\rm opt}_{{\rm p}, k}}})$ and the maximum SINR of UE $k$ is achieved at $b=b_k^{\rm opt}$. This completes the proof.


\bibliographystyle{IEEEtran}
\bibliography{IEEEabrv, Reference}

\end{document}